\documentclass[12pt]{article}
\usepackage[margin=0.5in]{geometry}
\usepackage{cancel}
\usepackage{amssymb}
\usepackage{amsmath}
\usepackage{verbatim}

\usepackage{mathrsfs}
\usepackage{color}
\usepackage{graphicx}
\usepackage{multicol}
\usepackage{enumerate}
\usepackage{setspace}
\usepackage{bigcircle}
\usepackage[round]{natbib}
\begin{document}
\title{\bf Wave turbulence in the two-layer ocean model} %and atmospheres
\author{Katie L Harper$^{1}$, Sergey V Nazarenko$^{1,2}$, Sergey B Medvedev$^{3}$ and Colm Connaughton$^{1,4,5}$}
\date{}
\maketitle
\bibliographystyle{abbrvnat}
%\doublespacing

$^{1}$ Mathematics Institute, University of Warwick, Gibbet Hill Road, Coventry, CV4 7AL, United Kingdom

$^{2}$ Laboratoire SPHYNX, Service de Physique de l'Etat Condense, DSM, IRAMIS, CEA, Saclay, CNRS URA 2464, 91191, Gif-sur-Yvette, France

$^{3}$ Institute of Computational Technologies SD RAS, Lavrentjev avenue 6, Novosibirsk, 630090, Russia

$^{4}$ Centre for Complexity Science, University of Warwick, Gibbet Hill Road, Coventry CV4 7AL, United Kingdom

$^{5}$ Okinawa Institute of Science and Technology Graduate University, 1919-1 Tancha Onna-son, Okinawa 904-0495, Japan

\begin{abstract}
This paper looks at the two-layer ocean model from a wave turbulence perspective. A symmetric form of the two-layer kinetic equation for Rossby waves is derived using canonical variables, allowing the turbulent cascade of energy between the barotropic and baroclinic modes to be studied. It turns out that energy is transferred via local triad interactions from the large-scale baroclinic modes to the baroclinic and barotropic modes at the Rossby deformation scale. From there it is then transferred to the large-scale barotropic modes via a nonlocal inverse transfer. Using scale separation a system of coupled equations were obtained for the small-scale baroclinic component and the large-scale barotropic component. Since the total energy of the small-scale component is not conserved, but the total barotropic plus baroclinic energy is conserved, the baroclinic energy loss at small scales will be compensated by the growth of the barotropic energy at large scales. It is found that this transfer is mostly anisotropic and mostly to the zonal component.
\end{abstract}

\section{Introduction}

In the ocean, due to the vast amount of processes operating at different temporal and spacial scales, direct use of the underlying Navier-Stokes equations are neither realistic nor feasible. Consequently, this makes simplified nonlinear models a very important tool for studying large-scale geophysical flows. A breakthrough in this modelling was achieved by Charney in 1948 \citep{charney1948}. He used the fact that large-scale geophysical flows are quasi-hydrostatic and quasi-adiabatic to derive his famous barotropic one-layer beta-plane equation. On the other hand, barotropic equations often do not adequately represent reality as they lack the mechanism for energy injection and generation of turbulence. The two-layer model first introduced by Phillips in 1951 \citep{phillips1951} represents the next level of realism in describing geophysical fluid dynamics (GFD). Indeed this model allows for baroclinic motions and as a result baroclinic instabilities which are believed to be the main source of energy for large-scale geophysical turbulence.

Similar to Kraichnan's 1967 \citep{kraichnan1967} picture of two-dimensional barotropic motion with the inverse energy and direct enstrophy cascades, there have been ideas put forward for two-layer baroclinic motion. Most notably Salmon \citep{salmon1978} suggested in 1978 that energy enters at large horizontal scales, $k_{F},$ in the baroclinic mode. It then moves towards higher wavenumbers until it reaches the Rossby deformation scale, $k_{R},$ (one over the Rossby deformation radius) where eddies generated through baroclinic instability (BI) \citep{lin1980} energise the barotropic mode. Energy then moves back towards large scales via an inverse barotropic cascade. In the presence of the beta-effect this inverse cascade is anisotropic meaning energy going to large scales will have a dominant zonal component. At scales smaller than the Rossby deformation scale there is a direct enstrophy cascade in each layer until it is scattered into three-dimensional turbulence.

The equations of geophysical fluid dynamics support large-scale wave motions known as Rossby waves. It has long been realised that triplets of Rossby wave modes known as resonant triads can play a special role in transferring  energy between scales. By a resonant triad we will understand a set of three modes for which {\em both} the wavenumbers and frequencies are in resonance i.e. $\mathbf{k}_{1}+\mathbf{k}_{2}=\mathbf{k}_{3}$ and $\omega_{\mathbf{k}_{1}}+\omega_{\mathbf{k}_{2}}=\omega_{\mathbf{k}_{3}}$ where $\omega_{\mathbf{k}}$ is the dispersion relation of the waves (see section \ref{sec: Two-layer equations}). Types of triad include - three barotropic modes, three baroclinic modes and those with mixed barotropic and baroclinic modes. Such resonant triads are effective when the level of geophysical turbulence is relatively weak. Otherwise non-resonant/vortex interactions, typical of non-rotating classical turbulence, would be more appropriate. Salmon \citep{salmon1978} studied the dynamics of the energy exchange within an individual non-resonant triad. However, in real geophysical turbulence many modes are excited simultaneously and therefore many coupled triads are active and mutually interacting. Such complex multi-dimensional systems call for a statistical description and this brings us to the domain of the wave turbulence (WT) approach. WT deals with correlators of the original wave fields, for the evolution of which the kinetic equation is derived \citep{zakharovlvovetal1992,nazarenko2011}.

WT is a tool that has proved valuable and effective in a great variety of cases from quantum to astrophysical scales i.e. quantum turbulence in superfluid helium \citep{lvovnazarenkoetal2006}, surface sea waves \citep{bedardlukaschuketal2013}, magneto-hydrodynamic turbulence  in astrophysics and laboratory plasmas \citep{tronkonazarenkoetal2013} and turbulence in rotating and stratified fluids \citep{bartello1995}. A great advantage of WT is that Kolmogorov-like spectra, for direct and inverse cascades, can be obtained as exact analytical solutions of the corresponding kinetic equations \citep{zakharovfilonenko1966}. They are called the Kolmogorov-Zakharov (KZ) spectra \citep{zakharovlvovetal1992,nazarenko2011}. Similar to Kolmogorov's solutions in classical hydrodynamic turbulence, the pure KZ solutions are only expected when the forcing and dissipation scales are separated by a wide inertial range of scales. In more realistic situations the kinetic equations remain useful in modelling both stationary and evolving turbulence spectra. For example, kinetic equations are widely used as a tool for day-to-day operational sea-wave weather forecasting \citep{janssen2008}.

For the two-layer baroclinic model the WT approach was first introduced by Kozlov et al. \citep{kozlovrezniketal1987} who derived kinetic equations for the barotropic and baroclinic Rossby wave components. They used a direct derivation with the physical variables for the normal barotropic and baroclinic modes. Unfortunately, the resulting kinetic equations were long and complicated when written in such variables, making them difficult to be used for further analysis. In the present paper we revisit WT theory for a two-layer ocean model. We derive a symmetric form of the Hamiltonian dynamical equations using the wave-action variables, which allows us to obtain kinetic equations which are compact and symmetric and therefore easier to use for modelling. Further, we use the kinetic equations to study the situation when the baroclinic and barotropic modes are scale separated. This gives us a very simple equation for the baroclinic energy spectrum which has the form of a diffusion equation in wavenumber space. We then use this equation to give a qualitative description of the coupled two component system, small-scale baroclinic waves and large-scale, zonally dominated barotropic turbulence.

The structure of this paper is as follows. In section \ref{sec: Two-layer equations}, we introduce the two-layer model and it's Hamiltonian form and following Kozlov et al. rewrite it in terms of the baroclinic and barotropic modes. In section \ref{sec: Fourier space}, we rewrite our equations in Fourier space and symmetrize them, under the assumption that the dominant interactions occur on the resonant manifold (i.e. such that both the wavenumbers and frequencies of the triads in the nonlinear term are in resonance). Such a symmetric Hamiltonian equation will serve as the starting point of the derivation of the kinetic equations, which is done in section \ref{sec: Derivation of the kinetic eqaution}. In section \ref{sec: Nonlocal interaction}, we derive a simplified diffusion equation for the scale separated system (small-scale baroclinic component and large-scale barotropic component) and build a qualitative picture of the evolution in such a system. In section \ref{sec: Summary}, we will present a summary of our results.

\section{Two-layer equations}\label{sec: Two-layer equations}

Let us consider two immiscible fluid layers, each with constant density $\rho_{i}$ and with the upper layer (i = 1) fluid lighter than the lower layer (i = 2) fluid, as required for gravitational
stability. The two-layer equations are:
\begin{eqnarray}
\frac{\partial}{\partial t}[\Delta\psi_{1}+\frac{f_{0}^{2}}{g'h_{1}}(\psi_{2}-\psi_{1})]+\beta\frac{\partial\psi_{1}}{\partial x}&=&-J[\psi_{1},\Delta\psi_{1}+\frac{f_{0}^{2}}{g'h_{1}}(\psi_{2}-\psi_{1})],\label{2-layer eqation a}\\
\frac{\partial}{\partial t}[\Delta\psi_{2}+\frac{f_{0}^{2}}{g'h_{2}}(\frac{\rho_{1}}{\rho_{2}}\psi_{1}-\psi_{2})]+\beta\frac{\partial\psi_{2}}{\partial x}&=&-J[\psi_{2},\Delta\psi_{2}+\frac{f_{0}^{2}}{g'h_{2}}(\frac{\rho_{1}}{\rho_{2}}\psi_{1}-\psi_{2})],\label{2-layer equation b}
\end{eqnarray}
where $\psi_{1}\equiv \psi_{1}(x,y,t), \psi_{2}\equiv \psi_{2}(x,y,t)$ are the stream functions for the upper and lower layers respectively, $J[a,b]=\partial_{x}a\partial_{y}b-\partial_{y}a\partial_{x}b$ is the Jacobian, $f_{0}$ is the mean value of the Coriolis parameter, $f=f_{0}+\beta,$ $g^{'}=g(\rho_{2}-\rho_{1}/\rho_{2})$ is the reduced gravity and $h_{i}$ is the height of each layer. For us to use the kinetic framework, these equations must first be modified so that each linear part contains only one unknown function. Doing this is equivalent to introducing normal modes:
\begin{equation}
\label{Normal modes}
\psi^{\sigma}=\psi_{1}+s^{\sigma}\psi_{2}, \quad \sigma=+,-
\end{equation}
where $\psi^{+}=\psi_{1}+s^{+}\psi_{2}$ is the barotropic normal mode and $\psi^{-}=\psi_{1}+s^{-}\psi_{2}$ is the baroclinic normal mode. Following the working of \citep{kozlovrezniketal1987} we get the single equation:
\begin{equation}
\label{Linear and nonlinear}
\frac{\partial}{\partial t}(\Delta \psi^{\sigma}-F^{\sigma}\psi^{\sigma})+\beta\frac{\partial \psi^{\sigma}}{\partial x}=-\lambda\sum\limits_{\mu \nu}[p_{\mu \nu}^{\sigma}J(\psi^{\mu},\Delta \psi^{\nu})+F^{\sigma}g_{\mu \nu}^{\sigma}J(\psi^{\mu},\psi^{\nu})],
\end{equation}
where $\lambda=(s^{+}-s^{-})$ and the coupling coefficients $p_{\mu \nu}^{\sigma}$ are as follows:
\begin{eqnarray}
\label{Coupling coefficients}
&&p_{++}^{+}=s^{+}+(s^{-})^{2}, \quad p_{--}^{+}=s^{+}(1+s^{+}), \quad p_{+-}^{+}=-s^{+}(1+s^{-})=p_{-+}^{+},\\ &&p_{++}^{-}=s^{-}(1+s^{-}), \quad p_{--}^{-}=s^{-}+(s^{+})^{2}, \quad p_{+-}^{-}=-s^{-}(1+s^{+})=p_{-+}^{-},\nonumber\\
&&\quad \quad \quad \quad \quad g_{+-}^{+}=-g_{-+}^{+}=-\frac{1}{2}(s^{+}-s^{-})=g_{+-}^{-}=-g_{-+}^{-}.\nonumber
\end{eqnarray}
\begin{equation}
F^{\sigma}=\frac{f_{0}^{2}}{g^{'}h_{1}}-\frac{\rho_{1}}{\rho_{2}}\frac{f_{0}^{2}}{g^{'}h_{2}}s^{\sigma},
\end{equation}
where $(F^{\sigma})^{-1/2}$ are the barotropic ($\sigma=+$) and the baroclinic ($\sigma=-$) Rossby deformation radii respectively. $s^{\pm}$ are solutions of the following quadratic equation:
\begin{equation}
\label{S}
\frac{\rho_{1}}{\rho_{2}}\frac{1}{h_{2}}s^{2}+(\frac{1}{h_{2}}-\frac{1}{h_{1}})s-\frac{1}{h_{1}}=0.
\end{equation}
Since the density of water masses in the Earth's oceans varies insignificantly, we have $(\rho_{2}-\rho_{1})/\rho_{2}<<1,$ so that $\rho_{1}/\rho_{2}\simeq 1.$ Consequently the roots of equation (\ref{S}) are:
\begin{equation}
s^{+}\simeq \frac{h_{2}}{h_{1}}, \quad s^{-}\simeq -1.
\end{equation}

\section{Fourier space representation}\label{sec: Fourier space}

Let us now put equation (\ref{Linear and nonlinear}) into Fourier space. Let the system be in a periodic box, with period $L$ in both directions. Fourier series representation of the barotropic and baroclinic stream function is:
\begin{equation}
\psi^{\sigma}(\mathbf{x},t)=\sum\limits_{\mathbf{k}} \hat{\psi}^{\sigma}(\mathbf{k},t)e^{i\mathbf{k}\cdot \mathbf{x}},
\end{equation}
with Fourier coefficients:
\begin{equation}
\hat{\psi}^{\sigma}(\mathbf{k},t)=\frac{1}{L^{2}}\int \psi^{\sigma}(\mathbf{x},t)e^{-i\mathbf{k}\cdot \mathbf{x}}d\mathbf{x},
\end{equation}
where $\mathbf{k}=(k_{x},k_{y})$ is a 2D wave vector taking values on a 2D discrete lattice: $k_{x}=2\pi l/L, k_{y}=2\pi m/L; l,m\in \mathbb{Z}^{2}.$
Fourier transforming equation (\ref{Linear and nonlinear}) we get:
\begin{eqnarray}
\frac{\partial}{\partial t}(-k^{2}-F^{\sigma})\hat{\psi}_{\mathbf{k}}^{\sigma}+i\beta k_{x}\hat{\psi}_{\mathbf{k}}^{\sigma}&=&-\lambda \sum\limits_{\mu \nu}\sum\limits_{\mathbf{k}_{1},\mathbf{k}_{2}}[p_{\mu \nu}^{\sigma}(k_{1x}\hat{\psi}_{1}^{\mu}k_{2y}k_{2}^{2}\hat{\psi}_{2}^{\nu}-k_{1y}\hat{\psi}_{1}^{\mu}k_{2x}k_{2}^{2}\hat{\psi}_{2}^{\nu})\\
&&+F^{\sigma}g_{\mu \nu}^{\sigma}(k_{1x}\hat{\psi}_{1}^{\mu}k_{2y}\hat{\psi}_{2}^{\nu}-k_{1y}\hat{\psi}_{1}^{\mu}k_{2x}\hat{\psi}_{2}^{\nu})]\delta_{12}^{\mathbf{k}},\nonumber
\end{eqnarray}
where $\hat{\psi}_{1}^{\mu}\equiv \hat{\psi}_{\mathbf{k}_{1}}^{\mu}$ and $\hat{\psi}_{2}^{\nu}\equiv \hat{\psi}_{\mathbf{k}_{2}}^{\nu}.$ Now dividing by $-(k^{2}+F^{\sigma})$ and swapping 1 and 2 we get:
\begin{equation}
\label{Fourier}
\partial_{t}\hat{\psi}_{\mathbf{k}}^{\sigma}+i\omega_{\mathbf{k}}^{\sigma}\hat{\psi}_{\mathbf{k}}^{\sigma}=\lambda \sum\limits_{\mu \nu}\sum\limits_{\mathbf{k}_{1},\mathbf{k}_{2}} D_{\mathbf{k}_{1}\mathbf{k}_{2}}^{\sigma \mu \nu}\hat{\psi}_{1}^{\mu}\hat{\psi}_{2}^{\nu}\delta_{12}^{\mathbf{k}},
\end{equation}
where:
\begin{equation}
\delta_{12}^{\mathbf{k}}=\delta(\mathbf{k}-\mathbf{k}_{1}-\mathbf{k}_{2}),
\end{equation}
\begin{equation}
\omega_{\mathbf{k}}^{\sigma}=-\frac{\beta k_{x}}{k^{2}+F^{\sigma}},
\end{equation}
\begin{equation}
D_{\mathbf{k}_{1}\mathbf{k}_{2}}^{\sigma \mu \nu}=\frac{C_{\mathbf{k}_{1}\mathbf{k}_{2}}^{\sigma \mu \nu}}{k^{2}+F^{\sigma}},
\end{equation}
and:
\begin{equation}
C_{\mathbf{k}_{1}\mathbf{k}_{2}}^{\sigma \mu \nu}=\frac{1}{2}p_{\mu \nu}^{\sigma}(\mathbf{k}_{1}\times \mathbf{k}_{2})_{z}(k_{2}^{2}-k_{1}^{2}+F^{\nu}-F^{\mu}).
\end{equation}

\subsection{Introducing the wave-action variable and symmetrization}

Let us introduce the wave-action variable:
\begin{equation}
a_{\mathbf{k}}^{\pm}=\frac{\hat{\psi}_{\mathbf{k}}^{\pm}}{\sqrt{|\omega_{\mathbf{k}}^{\pm}s^{\pm}|}}.
\end{equation}
Substituting this into equation (\ref{Fourier}) we get:
\begin{eqnarray}
\dot{a}_{\mathbf{k}}^{\sigma}\frac{\sqrt{|k_{x}|}\sqrt{|s^{\sigma}|}}{(k^{2}+F^{\sigma})}+i\omega_{\mathbf{k}}^{\sigma}a_{\mathbf{k}}^{\sigma}\frac{\sqrt{|k_{x}|}\sqrt{|s^{\sigma}|}}{(k^{2}+F^{\sigma})}&=&\frac{\lambda}{2(k^{2}+F^{\sigma})}\sum\limits_{\mu \nu}\sum\limits_{\mathbf{k}_{1},\mathbf{k}_{2}}p_{\mu \nu}^{\sigma}(k_{1x}k_{2y}-k_{2x}k_{1y})\\
&&\times (k_{2}^{2}-k_{1}^{2}+F^{\nu}-F^{\mu})\bar{a}_{1}^{\mu}\bar{a}_{2}^{\nu}\frac{\sqrt{|k_{1x}|}\sqrt{|k_{2x}|}\sqrt{|s^{\mu}|}\sqrt{|s^{\nu}|}}{(k_{1}^{2}+F^{\mu})(k_{2}^{2}+F^{\nu})}\delta_{\mathbf{k}12},\nonumber
\end{eqnarray}
where $a_{1}^{\mu}\equiv a_{\mathbf{k}_{1}}^{\mu}$, $a_{2}^{\nu}\equiv a_{\mathbf{k}_{2}}^{\nu}$ and $\dot{a}_{\mathbf{k}}^{\sigma}$ denotes the derivative. To make the equation more symmetric we rewrote the Kronecker delta, $\delta_{12}^{\mathbf{k}}=\delta(\mathbf{k}-\mathbf{k}_{1}-\mathbf{k}_{2})$ as $\delta_{\mathbf{k}12}=\delta(\mathbf{k}+\mathbf{k}_{1}+\mathbf{k}_{2}).$ In order to do this we changed $\mathbf{k}_{1}\rightarrow -\mathbf{k}_{1}$ and $\mathbf{k}_{2}\rightarrow -\mathbf{k}_{2}$. Consequently $a_{-\mathbf{k}_{1}}=\bar{a}_{\mathbf{k}_{1}}$ and $a_{-\mathbf{k}_{2}}=\bar{a}_{\mathbf{k}_{2}},$ where $\bar{a}$ denotes the complex conjugate. Rearranging the above equation we get:
\begin{eqnarray}
\dot{a}_{\mathbf{k}}^{\sigma}+i\omega_{\mathbf{k}}^{\sigma}a_{\mathbf{k}}^{\sigma}&=&-\frac{\lambda}{2}\sum\limits_{\mu \nu}\sum\limits_{\mathbf{k}_{1},\mathbf{k}_{2}}p_{\mu \nu}^{\sigma}(k_{x}k_{2y}-k_{2x}k_{y})\\
&&\times \frac{(k_{2}^{2}-k_{1}^{2}+F^{\nu}-F^{\mu})}{(k_{1}^{2}+F^{\mu})(k_{2}^{2}+F^{\nu})}\bar{a}_{1}^{\mu}\bar{a}_{2}^{\nu}\left|\frac{k_{1x}k_{2x}}{k_{x}}\right|^{1/2}\left|\frac{s^{\mu}s^{\nu}}{s^{\sigma}}\right|^{1/2}\delta_{\mathbf{k}12},\nonumber\\
&&\nonumber\\
&=&-\frac{\lambda}{2}\text{ sign}(k_{x})\sum\limits_{\mu \nu}\sum\limits_{\mathbf{k}_{1},\mathbf{k}_{2}}p_{\mu \nu}^{\sigma}|k_{1x}k_{2x}k_{x}|^{1/2}(k_{2y}-k_{2x}k_{y}/k_{x})\nonumber\\
&&\times \left(\frac{1}{k_{1}^{2}+F^{\mu}}-\frac{1}{k_{2}^{2}+F^{\nu}}\right)\bar{a}_{1}^{\mu}\bar{a}_{2}^{\nu}\left|\frac{s^{\mu}s^{\nu}}{s^{\sigma}}\right|^{1/2}\delta_{\mathbf{k}12},\nonumber\\
&&\nonumber\\
&=&-\frac{\lambda}{2}\text{ sign}(k_{x})\sum\limits_{\mu \nu}\sum\limits_{\mathbf{k}_{1},\mathbf{k}_{2}}p_{\mu \nu}^{\sigma}|k_{1x}k_{2x}k_{x}|^{1/2}\bar{a}_{1}^{\mu}\bar{a}_{2}^{\nu}\left|\frac{s^{\mu}s^{\nu}}{s^{\sigma}}\right|^{1/2}\nonumber\\
&&\times \left(\frac{k_{2y}}{k_{1}^{2}+F^{\mu}}-\frac{k_{2y}}{k_{2}^{2}+F^{\nu}}-\frac{k_{2x}k_{y}/k_{x}}{k_{1}^{2}+F^{\mu}}+\frac{k_{2x}k_{y}/k_{x}}{k_{2}^{2}+F^{\nu}}\right)\delta_{\mathbf{k}12}.\nonumber
\end{eqnarray}
Substituting the resonant conditions $k_{2y}=-k_{y}-k_{1y}$ and $k_{2x}=-k_{x}-k_{1x}$ into the brackets we get:
\begin{eqnarray}
\dot{a}_{\mathbf{k}}^{\sigma}+i\omega_{\mathbf{k}}^{\sigma}a_{\mathbf{k}}^{\sigma}&=&-\frac{\lambda}{2}\text{ sign}(k_{x})\sum\limits_{\mu \nu}\sum\limits_{\mathbf{k}_{1},\mathbf{k}_{2}}p_{\mu \nu}^{\sigma}|k_{1x}k_{2x}k_{x}|^{1/2}\bar{a}_{1}^{\mu}\bar{a}_{2}^{\nu}\left|\frac{s^{\mu}s^{\nu}}{s^{\sigma}}\right|^{1/2}\\
&&\times \left(\frac{-k_{y}-k_{1y}}{k_{1}^{2}+F^{\mu}}-\frac{k_{2y}}{k_{2}^{2}+F^{\nu}}-\frac{(-k_{x}-k_{1x})k_{y}/k_{x}}{k_{1}^{2}+F^{\mu}}+\frac{k_{2x}k_{y}/k_{x}}{k_{2}^{2}+F^{\nu}}\right)\delta_{\mathbf{k}12},\nonumber\\
&&\nonumber\\
&=&-\frac{\lambda}{2}\text{ sign}(k_{x})\sum\limits_{\mu \nu}\sum\limits_{\mathbf{k}_{1},\mathbf{k}_{2}}p_{\mu \nu}^{\sigma}|k_{1x}k_{2x}k_{x}|^{1/2}\bar{a}_{1}^{\mu}\bar{a}_{2}^{\nu}\left|\frac{s^{\mu}s^{\nu}}{s^{\sigma}}\right|^{1/2}\nonumber\\
&&\times \left(\frac{-k_{1y}}{k_{1}^{2}+F^{\mu}}-\frac{k_{2y}}{k_{2}^{2}+F^{\nu}}+\frac{k_{1x}k_{y}/k_{x}}{k_{1}^{2}+F^{\mu}}+\frac{k_{2x}k_{y}/k_{x}}{k_{2}^{2}+F^{\nu}}\right)\delta_{\mathbf{k}12}.\nonumber
\end{eqnarray}
Finally using the resonant condition $-\omega_{\mathbf{k}}^{\sigma}=\omega_{1}^{\mu}+\omega_{2}^{\nu}$ we get:
\begin{equation}
\label{In terms of a}
\dot{a}_{\mathbf{k}}^{\sigma}+i\omega_{\mathbf{k}}^{\sigma}a_{\mathbf{k}}^{\sigma}=\text{ sign}(k_{x})\sum\limits_{\mu \nu}\sum\limits_{\mathbf{k}_{1},\mathbf{k}_{2}}V_{\mathbf{k}12}^{\sigma \mu \nu}p_{\mu \nu}^{\sigma}\bar{a}_{1}^{\mu}\bar{a}_{2}^{\nu}\left|\frac{s^{\mu}s^{\nu}}{s^{\sigma}}\right|^{1/2}\delta_{\mathbf{k}12},
\end{equation}
where:
\begin{equation}
V_{\mathbf{k}12}^{\sigma \mu \nu}=\frac{\lambda}{2}|k_{1x}k_{2x}k_{x}|^{1/2}\left(\frac{k_{1y}}{k_{1}^{2}+F^{\mu}}+\frac{k_{2y}}{k_{2}^{2}+F^{\nu}}+\frac{k_{y}}{k^{2}+F^{\sigma}}\right),
\end{equation}
is the nonlinear interaction coefficient for the wave-action variable.

\section{Derivation of the kinetic equations}\label{sec: Derivation of the kinetic eqaution}

\subsection{Time-scale separation}\label{subsec: Time-scale separation}

Let us rewrite equation (\ref{In terms of a}) in terms of an interaction representation variable:
\begin{equation}
b_{\mathbf{k}}^{\pm}=a_{\mathbf{k}}^{\pm}e^{i\omega_{\mathbf{k}}^{\pm}t},
\end{equation}
to get:
\begin{equation}
\label{Interaction rep.}
i\dot{b}_{\mathbf{k}}^{\sigma}=\text{ sign}(k_{x})\sum\limits_{\mu \nu}\sum\limits_{12}V_{\mathbf{k}12}^{\sigma \mu \nu}p_{\mu \nu}^{\sigma}\bar{b}_{1}^{\mu}\bar{b}_{2}^{\nu}e^{i\omega_{\mathbf{k}12}^{\sigma \mu \nu}t}\left|\frac{s^{\mu}s^{\nu}}{s^{\sigma}}\right|^{1/2}\delta_{\mathbf{k}12},
\end{equation}
where $b_{1}^{\mu}\equiv b_{\mathbf{k}_{1}}^{\mu},$ $b_{2}^{\nu}\equiv b_{\mathbf{k}_{2}}^{\nu},$ $\sum\limits_{12}=\sum\limits_{\mathbf{k}_{1},\mathbf{k}_{2}}$ and $\omega_{\mathbf{k}12}^{\sigma \mu \nu}=\omega_{\mathbf{k}}^{\sigma}+\omega_{\mathbf{k}_{1}}^{\mu}+\omega_{\mathbf{k}_{2}}^{\nu}.$ Let us assume that the wave amplitudes are small and nonlinearity is weak and separate the linear and nonlinear time scales as follows:
\begin{equation}
\tau_{L}=\frac{2\pi}{\omega_{\mathbf{k}}}<<\tau_{NL}=\frac{2\pi}{\epsilon^{2}\omega_{\mathbf{k}}},
\end{equation}
in order to filter out the fast oscillatory motions and describe the slowly changing wave statistics. Let us now introduce an intermediate time $T=\frac{2\pi}{\epsilon\omega_{\mathbf{k}}}$ and find a solution for the wave amplitudes $b_{\mathbf{k}}^{\pm}$ at time $t=T$ using the following expansion in the small nonlinearity parameter $\epsilon \ll 1$ \citep{nazarenko2011}:
\begin{equation}
\label{Perturbation series}
b_{\mathbf{k}}^{\pm}(T)=b_{\mathbf{k}}^{\pm (0)}+\epsilon b_{\mathbf{k}}^{\pm (1)}+\epsilon^{2}b_{\mathbf{k}}^{\pm (2)}+...
\end{equation}
The first term in the expansion $\mathcal{O}(\epsilon^{0})$ corresponds to the linear approximation in which the interaction representation amplitude is time independent:
\begin{equation}
\label{Order 0}
b_{\mathbf{k}}^{\pm (0)}(T)=b_{\mathbf{k}}^{\pm}(0).
\end{equation}
Now substitute $b_{\mathbf{k}}^{\pm (0)}$ into the right hand side of equation (\ref{Interaction rep.}) to get $\mathcal{O}(\epsilon^{1})$:
\begin{equation}
i\dot{b}_{\mathbf{k}}^{\sigma (1)}=\text{ sign}(k_{x})\sum\limits_{\mu \nu}\sum\limits_{12}V_{\mathbf{k}12}^{\sigma \mu \nu}p_{\mu \nu}^{\sigma}\bar{b}_{1}^{\mu (0)}\bar{b}_{2}^{\nu (0)}e^{i\omega_{\mathbf{k}12}^{\sigma \mu \nu}t}\left|\frac{s^{\mu}s^{\nu}}{s^{\sigma}}\right|^{1/2}\delta_{\mathbf{k}12},
\end{equation}
and integrate to get:
\begin{equation}
\label{Order 1}
b_{\mathbf{k}}^{\sigma (1)}(T)=-i\text{ sign}(k_{x})\sum\limits_{\mu \nu}\sum\limits_{12}V_{\mathbf{k}12}^{\sigma \mu \nu}p_{\mu \nu}^{\sigma}\bar{b}_{1}^{\mu (0)}\bar{b}_{2}^{\nu (0)}\Delta_{T}(\omega_{\mathbf{k}12}^{\sigma \mu \nu})\left|\frac{s^{\mu}s^{\nu}}{s^{\sigma}}\right|^{1/2}\delta_{\mathbf{k}12},
\end{equation}
where:
\begin{equation}
\Delta_{T}(\omega_{\mathbf{k}12}^{\sigma \mu \nu})=\int\limits_{0}^{T}e^{i\omega_{\mathbf{k}12}^{\sigma \mu \nu}t}dt.
\end{equation}
For the $\mathcal{O}(\epsilon^{2})$ term we get:
\begin{equation}
\label{O(2)}
i\dot{b}_{\mathbf{k}}^{\sigma (2)}=2\text{ sign}(k_{x})\sum\limits_{\mu \nu}\sum\limits_{12}V_{\mathbf{k}12}^{\sigma \mu \nu}p_{\mu \nu}^{\sigma}\bar{b}_{1}^{\mu (1)}\bar{b}_{2}^{\nu (0)}e^{i\omega_{\mathbf{k}12}^{\sigma \mu \nu}t}\left|\frac{s^{\mu}s^{\nu}}{s^{\sigma}}\right|^{1/2}\delta_{\mathbf{k}12},
\end{equation}
where $2$ arises due to the symmetry with respect to changing indices $1\leftrightarrow 2.$ Substitute $b_{\mathbf{k}}^{\sigma (1)}$ from equation (\ref{Order 1}) into (\ref{O(2)}) and integrate to get:
\begin{equation}
\label{Order 2}
b_{\mathbf{k}}^{\sigma (2)}(T)=-2\text{ sign}(k_{x}k_{1x})\sum\limits_{\mu \nu}\sum\limits_{1234}V_{\mathbf{k}12}^{\sigma \mu \nu}V_{134}^{\nu \sigma \mu}p_{\mu \nu}^{\sigma}p_{\sigma \mu}^{\nu}|s^{\mu}|\bar{b}_{2}^{\nu (0)}b_{3}^{\mu (0)}b_{4}^{\nu (0)}E(\omega_{134}^{\nu \sigma \mu},\omega_{\mathbf{k}12}^{\sigma \mu \nu})\delta_{\mathbf{k}12}\delta_{134},
\end{equation}
where:
\begin{equation}
E(\omega_{134}^{\nu \sigma \mu},\omega_{\mathbf{k}12}^{\sigma \mu \nu})=\int\limits_{0}^{T}\Delta_{T}(\omega_{134}^{\nu \sigma \mu})e^{i\omega_{\mathbf{k}12}^{\sigma \mu \nu}t}dt,
\end{equation}
and $\sum\limits_{1234}\equiv \sum\limits_{\mathbf{k}_{1},\mathbf{k}_{2},\mathbf{k}_{3},\mathbf{k}_{4}}.$ We do not need to find higher-order terms in the expansion since a non-trivial closure arises in the order $\epsilon^{2}.$

\subsection{Statistical averaging}\label{subsec: Statistical averaging}

The dynamical equations above describe the time evolution of wave amplitudes and phases. At weak nonlinearity with a large number of excited waves such a description is generally redundant so the dynamical description of a wave system is replaced by a statistical one in terms of correlation functions of the field \citep{nazarenko2011}. First we must do a weak nonlinearity expansion for the one-mode generating function at the intermediate time T as follows:
\begin{eqnarray}
\label{Statistical average}
<|b_{\mathbf{k}}^{\pm}(T)|^{2}>&=&<|b_{\mathbf{k}}^{\pm (0)}+\epsilon b_{\mathbf{k}}^{\pm (1)}+\epsilon^{2}b_{\mathbf{k}}^{\pm (2)}|^{2}>\\
&=&<|b_{\mathbf{k}}^{\pm (0)}|^{2}+\epsilon(|\bar{b}_{\mathbf{k}}^{\pm (0)}b_{\mathbf{k}}^{\pm (1)}|+c.c.)+\epsilon^{2}|b_{\mathbf{k}}^{\pm (1)}|^{2}+\epsilon^{2}(|\bar{b}_{\mathbf{k}}^{\pm (0)}b_{\mathbf{k}}^{\pm (2)}|+c.c.)>,\nonumber
\end{eqnarray}
where $<>$ denotes the ensemble average. Now we can perform statistical averaging over the random phases and amplitudes, starting with the former. The $\epsilon^{1}$ term using equation (\ref{Order 1}) is:
\begin{eqnarray}
\label{e1}
<|\bar{b}_{\mathbf{k}}^{\pm (0)}b_{\mathbf{k}}^{\pm (1)}|>_{\varphi}&=&-i\text{ sign}(k_{x})\sum\limits_{\mu \nu}\sum\limits_{\mathbf{k}12}V_{\mathbf{k}12}^{\sigma \mu \nu}p_{\mu \nu}^{\sigma}\left|\frac{s^{\mu}s^{\nu}}{s^{\sigma}}\right|^{1/2}<\bar{b}_{\mathbf{k}}^{\pm (0)}\bar{b}_{1}^{\mu (0)}\bar{b}_{2}^{\nu (0)}>_{\varphi}\Delta_{T}(\omega_{\mathbf{k}12}^{\sigma \mu \nu})\delta_{\mathbf{k}12}+c.c.
\end{eqnarray}
Wick's contraction rule \citep{nazarenko2011} states that $<\psi_{l1},\psi_{l2},...,\bar{\psi}_{m1},\bar{\psi}_{m2}>$ is zero unless the number of $\psi$'s in it equal the number of $\bar{\psi}$'s. So by Wick's contraction rule, since the correlation function in equation (\ref{e1}) has an odd number of terms, it and its complex conjugate are zero. The first $\epsilon^{2}$ term is:
\begin{eqnarray}
\label{e2.1}
<|b_{\mathbf{k}}^{\pm (1)}|^{2}>_{\varphi}&=&\sum\limits_{\mu \nu}\sum\limits_{1234}V_{\mathbf{k}12}^{\sigma \mu \nu}\bar{V}_{\mathbf{k}34}^{\sigma \mu \nu}(p_{\mu \nu}^{\sigma})^{2}\left|\frac{s^{\mu}s^{\nu}}{s^{\sigma}}\right|<b_{1}^{\mu (0)}b_{2}^{\nu (0)}\bar{b}_{3}^{\mu (0)}\bar{b}_{4}^{\nu (0)}>_{\varphi}\\
&&\times \Delta_{T}(\omega_{\mathbf{k}12}^{\sigma \mu \nu})\bar{\Delta_{T}}(\omega_{\mathbf{k}34}^{\sigma \mu \nu})\delta_{\mathbf{k}12}\delta_{\mathbf{k}34}.\nonumber
\end{eqnarray}
In equation (\ref{e2.1}) we must look for combinations of wave vectors in the fourth order correlator\\ 
$<b_{1}^{\mu (0)}b_{2}^{\nu (0)}\bar{b}_{3}^{\mu (0)}\bar{b}_{4}^{\nu (0)}>_{\varphi}$ that give a nonzero phase average. Replacing the four-point function by a product of two two-point functions we have \citep{nazarenko2011}:
\begin{eqnarray}
<b_{1}^{\mu (0)}b_{2}^{\nu (0)}\bar{b}_{3}^{\mu (0)}\bar{b}_{4}^{\nu (0)}>&=&<b_{1}^{\mu (0)}\bar{b}_{3}^{\mu (0)}><b_{2}^{\nu (0)}\bar{b}_{4}^{\nu (0)}>+<b_{1}^{\mu (0)}\bar{b}_{4}^{\nu (0)}><b_{2}^{\nu (0)}\bar{b}_{3}^{\mu (0)}>\\
&&+<b_{1}^{\mu (0)}\bar{b}_{-2}^{\nu (0)}><\bar{b}_{3}^{\mu (0)}b_{-4}^{\nu (0)}>,\nonumber
\end{eqnarray}
i.e. wavenumbers $\mathbf{k}_{1}=\mathbf{k}_{3}$ and $\mathbf{k}_{2}=\mathbf{k}_{4}$ or $\mathbf{k}_{1}=\mathbf{k}_{4}$ and $\mathbf{k}_{2}=\mathbf{k}_{3}$. Since $\bar{b}_{\mathbf{k}}=b_{-\mathbf{k}}$ we also have $\mathbf{k}_{1}=-\mathbf{k}_{2}$ and $\mathbf{k}_{3}=-\mathbf{k}_{4}$. The first two are the same from the $1\leftrightarrow 2$ symmetry. The last combination has zero delta's because $\mathbf{k}_{1}=-\mathbf{k}_{2} \Rightarrow \mathbf{k}-\mathbf{k}_{2}+\mathbf{k}_{2}=0$ which is impossible since $\mathbf{k}\neq 0$ and similarly for $\mathbf{k}_{3}=-\mathbf{k}_{4}$.

Defining:
\begin{equation}
J_{k}^{\pm}=|b_{\mathbf{k}}^{\pm}|^{2},
\end{equation}
equation (\ref{e2.1}) becomes:
\begin{equation}
\label{2.1}
<|b_{\mathbf{k}}^{\pm (1)}|^{2}>_{\varphi}=2\sum\limits_{\mu \nu}\sum\limits_{12}|V_{\mathbf{k}12}^{\sigma \mu \nu}|^{2}(p_{\mu \nu}^{\sigma})^{2}\left|\frac{s^{\mu}s^{\nu}}{s^{\sigma}}\right|J_{1}^{\mu}J_{2}^{\nu}|\Delta_{T}(\omega_{\mathbf{k}12}^{\sigma \mu \nu})|^{2}\delta_{\mathbf{k}12}.
\end{equation}
The second $\epsilon^{2}$ term using equation (\ref{Order 2}) is:
\begin{eqnarray}
\label{e2.2}
<|\bar{b}_{\mathbf{k}}^{\pm (0)}b_{\mathbf{k}}^{\pm (2)}|>_{\varphi}&=&-2\text{ sign}(k_{x})\text{ sign}(k_{1x})\sum\limits_{\mu \nu}\sum\limits_{1234}V_{\mathbf{k}12}^{\sigma \mu \nu}V_{134}^{\nu \sigma \mu}p_{\mu \nu}^{\sigma}p_{\sigma \mu}^{\nu}|s^{\mu}|<\bar{b}_{\mathbf{k}}^{\pm (0)}\bar{b}_{2}^{\nu (0)}b_{3}^{\mu (0)}b_{4}^{\nu (0)}>_{\varphi}\\
&&\times E(\omega_{134}^{\nu \sigma \mu},\omega_{\mathbf{k}12}^{\sigma \mu \nu})\delta_{\mathbf{k}12}\delta_{134}+c.c.\nonumber
\end{eqnarray}
Again we look for combinations of wave vectors that give a nonzero phase average. We have $\mathbf{k}=\mathbf{k}_{3}$ and $\mathbf{k}_{2}=\mathbf{k}_{4}$ or $\mathbf{k}=\mathbf{k}_{4}$ and $\mathbf{k}_{2}=\mathbf{k}_{3}$ or $\mathbf{k}=-\mathbf{k}_{2}$ and $\mathbf{k}_{3}=-\mathbf{k}_{4}$. The first two are the same from the $3\leftrightarrow 4$ symmetry. The last combination is ruled out as the deltas are zero. So equation (\ref{e2.2}) becomes:
\begin{equation}
\label{2.2}
<|\bar{b}_{\mathbf{k}}^{\pm (0)}b_{\mathbf{k}}^{\pm (2)}|>_{\varphi}=-4\text{ sign}(k_{x}k_{1x})\sum\limits_{\mu \nu}\sum\limits_{12}|V_{\mathbf{k}12}^{\sigma \mu \nu}|^{2}p_{\mu \nu}^{\sigma}p_{\sigma \mu}^{\nu}|s^{\mu}|J_{\mathbf{k}}^{\sigma}J_{2}^{\nu}E(\omega_{134}^{\nu \sigma \mu},\omega_{\mathbf{k}12}^{\sigma \mu \nu})\delta_{\mathbf{k}12}.
\end{equation}
We now want to perform amplitude averaging and introduce the wave spectrum:
\begin{equation}
<J_{\mathbf{k}}^{\pm}>=(\frac{2\pi}{L})^{2}n_{\mathbf{k}}^{\pm},
\end{equation}
into equations (\ref{2.1}) and (\ref{2.2}). Summing the resulting two equations we get:
%A change in the wave-action spectrum, $n_{\mathbf{k}},$ can be determined by the change in the probability %amplitude, given by:
%\begin{equation}
%\delta n_{\mathbf{k}}=|b_{\mathbf{k}}^{\pm}(T)|^{2}-|b_{\mathbf{k}}^{\pm}(0)|^{2}.
%\end{equation}
\begin{eqnarray}
(\frac{L}{2\pi})^{2}<|b_{\mathbf{k}}^{\pm}(T)|^{2}-|b_{\mathbf{k}}^{\pm}(0)|^{2}>&=&2(\frac{2\pi}{L})^{2}\sum\limits_{\mu \nu}\sum\limits_{12}|V_{\mathbf{k}12}^{\sigma \mu \nu}|^{2}(p_{\mu \nu}^{\sigma})^{2}\left|\frac{s^{\mu}s^{\nu}}{s^{\sigma}}\right|n_{1}^{\mu}n_{2}^{\nu}|\Delta_{T}(\omega_{\mathbf{k}12}^{\sigma \mu \nu})|^{2}\delta_{\mathbf{k}12}\\
&&+4(\frac{2\pi}{L})^{2}\text{ sign}(k_{x}k_{1x})\sum\limits_{\mu \nu}\sum\limits_{12}|V_{\mathbf{k}12}^{\sigma \mu \nu}|^{2}p_{\mu \nu}^{\sigma}p_{\sigma \mu}^{\nu}|s^{\mu}|n_{\mathbf{k}}^{\sigma}n_{2}^{\nu}E(\omega_{134}^{\nu \sigma \mu},\omega_{\mathbf{k}12}^{\sigma \mu \nu})\delta_{\mathbf{k}12},\nonumber
\end{eqnarray}
where $n_{1}^{\mu}\equiv n_{\mathbf{k}_{1}}^{\mu}$ and $n_{2}^{\nu}\equiv n_{\mathbf{k}_{2}}^{\nu}.$ In the large box limit, taking $L \rightarrow \infty$ we get:
\begin{eqnarray}
\label{Large box limit}
(\frac{L}{2\pi})^{2}<|b_{\mathbf{k}}^{\pm}(T)|^{2}-|b_{\mathbf{k}}^{\pm}(0)|^{2}>&=&2\sum\limits_{\mu \nu}\int|V_{\mathbf{k}12}^{\sigma \mu \nu}|^{2}(p_{\mu \nu}^{\sigma})^{2}\left|\frac{s^{\mu}s^{\nu}}{s^{\sigma}}\right|n_{1}^{\mu}n_{2}^{\nu}|\Delta_{T}(\omega_{\mathbf{k}12}^{\sigma \mu \nu})|^{2}\delta_{\mathbf{k}12}d\mathbf{k}_{12}\\
&&+4\text{ sign}(k_{x}k_{1x})\sum\limits_{\mu \nu}\int|V_{\mathbf{k}12}^{\sigma \mu \nu}|^{2}p_{\mu \nu}^{\sigma}p_{\sigma \mu}^{\nu}|s^{\mu}|n_{\mathbf{k}}^{\sigma}n_{2}^{\nu}E(\omega_{134}^{\nu \sigma \mu},\omega_{\mathbf{k}12}^{\sigma \mu \nu})\delta_{\mathbf{k}12}d\mathbf{k}_{12}.\nonumber
\end{eqnarray}
Taking the weak nonlinearity limit, $\epsilon \rightarrow 0,$ in the intermediate time $T=2\pi/\epsilon \omega_{\mathbf{k}}$ we get $T\rightarrow \infty$ \citep{nazarenko2011}. In equation (\ref{Large box limit}) we have:
\begin{equation}
|\Delta_{T}(\omega_{\mathbf{k}12}^{\sigma \mu \nu})|^{2}\rightarrow 2\pi T \delta (\omega_{\mathbf{k}12}^{\sigma \mu \nu}),
\end{equation}
and:
\begin{equation}
E(\omega_{134}^{\nu \sigma \mu},\omega_{\mathbf{k}12}^{\sigma \mu \nu})\rightarrow 2\pi T \delta(\omega_{\mathbf{k}12}^{\sigma \mu \nu}).
\end{equation}
Defining:
\begin{equation}
\frac{\partial n_{\mathbf{k}}^{\pm}}{\partial t}\simeq (\frac{L}{2\pi})^{2}\frac{<|b_{\mathbf{k}}^{\pm}(T)|^{2}-|b_{\mathbf{k}}^{\pm}(0)|^{2}>}{T},
\end{equation}
we get:
\begin{eqnarray}
\frac{\partial n_{\mathbf{k}}^{\pm}}{\partial t}&=&2\sum\limits_{\mu \nu}\int|V_{\mathbf{k}12}^{\sigma \mu \nu}|^{2}(p_{\mu \nu}^{\sigma})^{2}\left|\frac{s^{\mu}s^{\nu}}{s^{\sigma}}\right|n_{1}^{\mu}n_{2}^{\nu}2\pi \delta (\omega_{\mathbf{k}12}^{\sigma \mu \nu})\delta_{\mathbf{k}12}d\mathbf{k}_{12}\\
&&+4\text{ sign}(k_{x}k_{1x})\sum\limits_{\mu \nu}\int|V_{\mathbf{k}12}^{\sigma \mu \nu}|^{2}p_{\mu \nu}^{\sigma}p_{\sigma \mu}^{\nu}|s^{\mu}|n_{\mathbf{k}}^{\sigma}n_{2}^{\nu}2\pi \delta(\omega_{\mathbf{k}12}^{\sigma \mu \nu})\delta_{\mathbf{k}12}d\mathbf{k}_{12}.\nonumber
\end{eqnarray}
And so we have the two-layer kinetic equation:
\begin{equation}
\label{2-layer kinetic equation}
\frac{\partial n_{\mathbf{k}}^{\sigma}}{\partial t}=4 \pi\sum\limits_{\mu \nu}\int |V_{\mathbf{k}12}^{\sigma \mu \nu}|^{2}\left[(p_{\mu \nu}^{\sigma})^{2}\left|\frac{s^{\mu}s^{\nu}}{s^{\sigma}}\right|n_{1}^{\mu}n_{2}^{\nu}+2p_{\mu \nu}^{\sigma}p_{\sigma\mu}^{\nu}|s^{\mu}|n_{\mathbf{k}}^{\sigma}n_{1}^{\mu}\text{sign}(k_{x}k_{2x})\right]\delta(\omega_{\mathbf{k}12}^{\sigma \mu \nu})\delta_{\mathbf{k}12}d\mathbf{k}_{12}.
\end{equation}
Now take the 8 different combinations of $\sigma,\mu,\nu=\pm$ i.e. $\{++-\}, \{-++\}, \{--+\}, \{+--\}, \{+-+\}, \{-+-\}, \{+++\}, \{---\}$ and substitute the coupling coefficients $p_{\mu \nu}^{\sigma}$ from (\ref{Coupling coefficients}) into equation (\ref{2-layer kinetic equation}). We finally get the two-layer kinetic equation in symmetric form:
\begin{equation}
\label{Symmetric 2-layer kinetic equation}
\frac{\partial n_{\mathbf{k}}^{\sigma}}{\partial t}=\sum\limits_{\mu \nu}\int W_{\mathbf{k}12}^{\sigma \mu \nu}n_{1}^{\mu}[n_{2}^{\nu}+2n_{\mathbf{k}}^{\sigma}\text{sign}(k_{x}k_{2x})]\delta(\omega_{\mathbf{k}12}^{\sigma \mu \nu})\delta_{\mathbf{k}12}d\mathbf{k}_{12}.
\end{equation}
Here:
\begin{equation}
W_{\mathbf{k}12}^{++-}=4\pi|V_{\mathbf{k}12}^{++-}|^{2}(1+s^{-})^{2}(s^{+})^{2}s^{-},\label{W++-}
\end{equation}
and the same for permutations of $+,+$ and $-$.\\
Symmetrically,
\begin{equation}
W_{\mathbf{k}12}^{--+}=4\pi|V_{\mathbf{k}12}^{--+}|^{2}(1+s^{+})^{2}(s^{-})^{2}s^{+},
\end{equation}
and the same for permutations of $-,-$ and $+$.
\begin{equation}
W_{\mathbf{k}12}^{+++}=4\pi|V_{\mathbf{k}12}^{+++}|^{2}(s^{+}+(s^{-})^2)s^{+},
\end{equation}
and symmetrically,
\begin{equation}
W_{\mathbf{k}12}^{---}=4\pi|V_{\mathbf{k}12}^{---}|^{2}(s^{-}+(s^{+})^2)s^{-}.\label{W---}
\end{equation}
The two-layer kinetic equations (\ref{Symmetric 2-layer kinetic equation}) conserves the energy:
\begin{equation}
E=\sum\limits_{\sigma}\int |\omega_{\mathbf{k}}^{\sigma}|n_{\mathbf{k}}^{\sigma}d\mathbf{k},
\end{equation}
and the potential enstrophy:
\begin{equation}
\Omega=\sum\limits_{\sigma}\int \omega_{\mathbf{k}}^{\sigma}(k^{2}+F^{\sigma})n_{\mathbf{k}}^{\sigma}d\mathbf{k}=\sum\limits_{\sigma}\int |k_{x}|n_{\mathbf{k}}^{\sigma}d\mathbf{k}.
\end{equation}
For the proof of this see appendix \ref{app: Conservation}.

As mentioned earlier, in the ocean the two layers will have different densities with the upper layer being only slightly lighter than the layer below. They will also have a significant difference in height, $h_{1}\ll h_{2},$ (typical ratio 1:7) where the subscript ``1'' corresponds to the upper layer and ``2'' to the lower layer. As a result we have: $$s^{-}\simeq -1,$$ and: $$s^{+}\simeq \frac{h_{2}}{h_{1}}\gg 1.$$
Substituting $s^{-}$ into $W_{\mathbf{k}12}^{++-}$ we can see that it vanishes. Since $s^{+}\gg 1$, $W_{\mathbf{k}12}^{--+}$ is the most dominant term since it contains an $(s^{+})^{3}$ term and $W_{\mathbf{k}12}^{+++}$ and $W_{\mathbf{k}12}^{---}$ only contain $(s^{+})^{2}$ terms. Hence we are left with permutations of $\{--+\}$ only, so the two-layer kinetic equations (\ref{Symmetric 2-layer kinetic equation}) reduces to:
\begin{eqnarray}
\label{Barotropic part}
\partial_{t}n_{\mathbf{k}}^{+}&=&\int W_{\mathbf{k}12}^{+--}[n_{1}^{-}n_{2}^{-}+2n_{1}^{-}n_{\mathbf{k}}^{+}sign(k_{x}k_{2x})]\delta(\omega_{\mathbf{k}}^{+}+\omega_{\mathbf{k}_{1}}^{-}+\omega_{\mathbf{k}_{2}}^{-})\delta(\mathbf{k}+\mathbf{k}_{1}+\mathbf{k}_{2})d\mathbf{k}_{12},\\
&&\nonumber\\
\label{Baroclinic part}
\partial_{t}n_{\mathbf{k}}^{-}&=&\int W_{\mathbf{k}12}^{-+-}[n_{1}^{+}n_{2}^{-}+n_{1}^{+}n_{\mathbf{k}}^{-}sign(k_{x}k_{2x})+n_{2}^{-}n_{\mathbf{k}}^{-}sign(k_{x}k_{1x})]\delta(\omega_{\mathbf{k}}^{-}+\omega_{\mathbf{k}_{1}}^{+}+\omega_{\mathbf{k}_{2}}^{-})\\
&&\delta(\mathbf{k}+\mathbf{k}_{1}+\mathbf{k}_{2})d\mathbf{k}_{12}.\nonumber
\end{eqnarray}

$\{--+\}$ is a triad with two baroclinic components and one barotropic component. In the barotropic part of the kinetic equation (\ref{Barotropic part}), $\mathbf{k}$ is the barotropic wavenumber and $\mathbf{k}_{1},\mathbf{k}_{2}$ are the baroclinic wavenumbers. In the baroclinic part (\ref{Baroclinic part}), $\mathbf{k}_{1}$ is the barotropic wavenumber and $\mathbf{k},\mathbf{k}_{2}$ are the baroclinic wavenumbers. Equally, we could have $W_{\mathbf{k}12}^{--+}$ in (\ref{Baroclinic part}) so that $\mathbf{k}_{2}$ is then the barotropic wavenumber and $\mathbf{k},\mathbf{k}_{1}$ are the baroclinic wavenumbers.

Since in the model of the ocean we are considering (a thin upper layer) $\{--+\}$ is the most dominant of the three triads we will focus on it entirely from now on. In the next section we will consider how energy is transferred between the baroclinic and the barotropic modes.

\section{Nonlocal interaction between baroclinic and barotropic modes}\label{sec: Nonlocal interaction}

\subsection{Energy transfer in two layers}\label{subsec: Energy transfer}

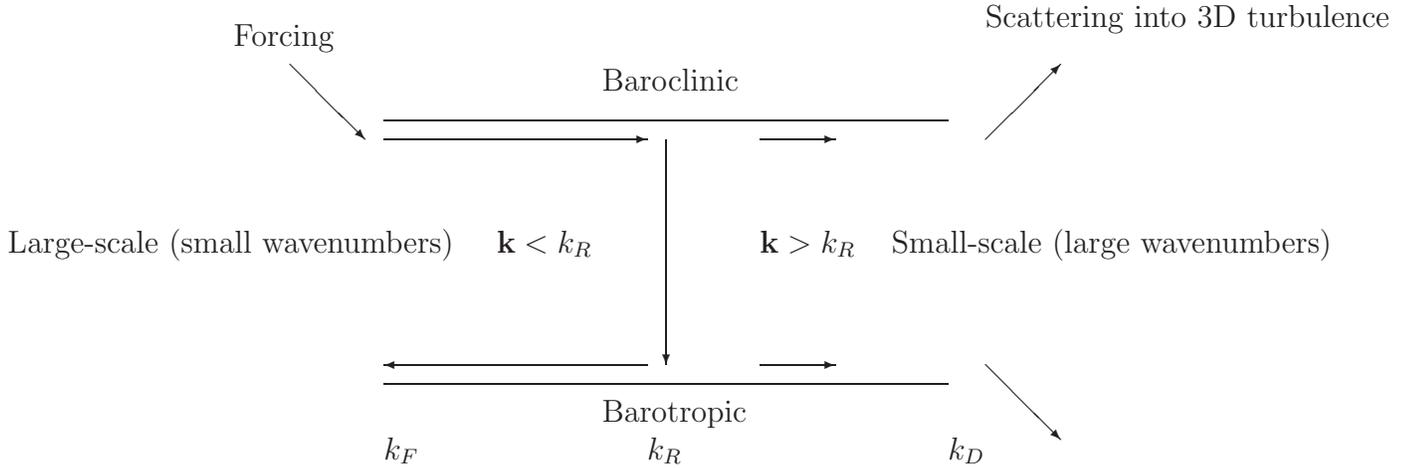
\begin{figure}[h!]
\centering
\setlength{\unitlength}{0.5mm}
\begin{picture}(150, 125)(0, -20)
  \put(0, 0){\line(1, 0){150}}
  \put(0, 70){\line(1, 0){150}}
  \put(0, 65){\vector(1, 0){70}}
  \put(75, 65){\vector(0, -1){60}}
  \put(100, 65){\vector(1, 0){20}}
  \put(100, 5){\vector(1, 0){20}}
  \put(70, 5){\vector(-1, 0){70}}
  \put(70, -20){$k_{R}$}
  \put(0,-20){$k_{F}$}
  \put(150, -20){$k_{D}$}
  \put(45, 30){}
  \put(58, 78){Baroclinic}
  \put(58, -10){Barotropic}
  \put(-100, 35){Large-scale (small wavenumbers)}
  \put(135, 35){Small-scale (large wavenumbers)}  
  \put(-25, 85){\vector(1, -1){20}}
  \put(-40, 90){Forcing}
  \put(160, 65){\vector(1, 1){20}}
  \put(160, 95){Scattering into 3D turbulence}
  \put(160, 5){\vector(1, -1){20}}
  \put(30, 35){$\mathbf{k}<k_{R}$}
  \put(100, 35){$\mathbf{k}>k_{R}$}
\end{picture}\\
\caption{Salmon's energy flux diagram for a two-layer system. The potential enstrophy flux present on the original diagram is omitted from our discussion.}
\label{fig: Salmon's picture}
\end{figure}

In the introduction we mentioned a schematic construction of energy flow in two layers which was first suggested by Salmon in 1978 \citep{salmon1978} and has since become the standard picture in geophysical literature. This picture is summarized in Salmon's diagram reproduced in figure \ref{fig: Salmon's picture}. It is important to realise that in his work Salmon considered equivalent layers i.e. equal depth and equal density. This means that $s^{+}=1$ and $s^{-}=-1$ and as a consequence only two types of triads $\{--+\}$ and $\{+++\}$ can exist (this can be seen from equations (\ref{W++-}) to (\ref{W---})).

In Salmon's picture energy is injected at the largest scale, $k_{F},$ via wind created by a temperature difference between the poles and the equator. Baroclinic modes then transfer this energy via nonlocal $\{--+\}$ triad interactions to the baroclinic and barotropic modes at the Rossby deformation scale, $k_{R}$. A small proportion of this energy will continue to flow to the smallest scale, $k_{D},$ where it is scattered into three-dimensional turbulence. However, the majority of the energy will be transferred to large-scale barotropic motions via local $\{+++\}$ triad interactions.

In this paper we will not consider scales less than the Rossby deformation scale, $\mathbf{k}>k_{R}$. Instead, we will concentrate on the energy transfer loop whereby the majority of the energy is transferred from the large-scale baroclinic mode to the large-scale barotropic mode in two steps:
\begin{enumerate}
\item The energy is transferred from the large-scale baroclinic modes to the baroclinic and barotropic modes at the Rossby deformation scale, $\mathbf{k}\sim k_{R}$.
\item It is then transferred from the baroclinic and barotropic modes at the Rossby deformation scale to the large-scale barotropic modes.
\end{enumerate}

A frequently discussed candidate mechanism for step 1 of this loop is baroclinic instability (BI). In Salmons paper \citep{salmon1978} BI is associated with a nonlocal triad interaction of the type $\{--+\}.$ In such a triad one baroclinic wavenumber is close to zero and unstable and will transfer energy nonlocally to the other baroclinic and barotropic wavenumbers with scales of the order of the Rossby deformation scale (i.e. $\mathbf{k}_{1}\approx \mathbf{k}_{2}>>\mathbf{k}\rightarrow 0$). This is more general than classical BI with a vertical shear which is the $\mathbf{k}=0$ limit. However, it can be shown that BI can not coexist with WT.

WT is applicable when the nonlinear terms in an equation are much less than the linear terms. In the case of the two-layer equations this is when:
\begin{equation}
J[\psi,\nabla^{2}\psi]\ll \beta\partial_{x}\psi.
\end{equation}
Let $k_{x}\sim k_{y}\sim\partial_{x}\sim\partial_{y}$ so we have:
\begin{equation}
\partial_{x}\psi\partial_{y}\nabla^{2}\psi\ll \beta\partial_{x}\psi \Rightarrow Uk^{2}\ll \beta.
\end{equation}
The necessary condition for BI is \citep{mcwilliams2006}:
\begin{equation}
U>\beta\rho^{2}.
\end{equation}
Putting the two conditions together we get:
\begin{equation}
\beta\rho^{2}k^{2}<Uk^{2}<\beta \Rightarrow \rho^{2}k^{2}\ll 1,
\end{equation}
but BI is maximum at $k_{R}$ (the scale of the Rossby deformation radius) and this is not described by WT. Thus we can conclude that WT and BI cannot operate simultaneously.

Let us also consider the possibility of nonlocal interaction from a WT perspective. The frequency resonance condition, $\omega_{\mathbf{k}}^{+}+\omega_{\mathbf{k}_{1}}^{-}+\omega_{\mathbf{k}_{2}}^{-}=0$ (which was not considered in Salmon's paper) must be satisfied along with the wavenumber resonance condition, $\mathbf{k}+\mathbf{k}_{1}+\mathbf{k}_{2}=0$. If $\mathbf{k}_{1}$ is small then $\mathbf{k}_{2}\approxeq -\mathbf{k}$ and the frequency condition gives $\omega_{\mathbf{k}}^{+}=\omega_{\mathbf{k}}^{-}$ which cannot be true. Hence, in our case the transfer of energy from the large-scale baroclinic modes to the Rossby deformation scale cannot be nonlocal and instead must be local. We will not consider this any further in this paper.

At step 2 of the loop, the energy accumulated at the Rossby deformation scale will be transferred into large-scale barotropic modes via an inverse transfer. This is similar to the one-layer case \citep{nazarenko2011,balknazarenkoetal1991,nazarenkoquinn2009} whereby the inverse energy transfer becomes anisotropic, with dominant zonal scales, due to the presence of a third invariant, zonostrophy. During this second stage the inverse cascade is most probably nonlocal. It may start off as a local cascade but will eventually lead to formation of strong (interaction with) zonal jets which will become dominant for $\mathbf{k}\sim k_{R}$ modes. In our two-layer system such one-layer interactions would correspond to the $\{+++\}$ triads. Drawing intuition from the one-layer case, it would be natural to assume that in our two-layer system the inverse energy transfer to the barotropic mode is also nonlocal. This is what we will now consider using a similar scale separation technique to that used for the one-layer model \citep{balknazarenkoetal1990,connaughtonnazarenkoetal2010}, but now for dominant $\{--+\}$ triads.

\subsection{Scale separation and the diffusion equation}\label{subsec: Scale separation}

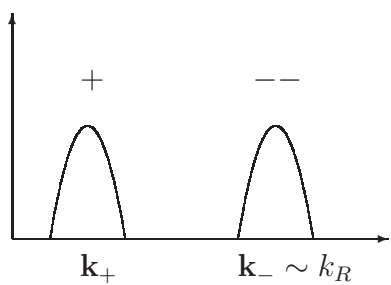
\begin{figure}[h!]
\centering
\setlength{\unitlength}{1mm}
\begin{picture}(60, 50)(0, 0)
\put(0, 20){\vector(1, 0){50}}
\put(0, 20){\vector(0, 1){30}}
\qbezier(5, 20)(10, 50)(15, 20)
\qbezier(30, 20)(35, 50)(40, 20)
\put(9, 15){$\mathbf{k}_{+}$}
\put(30,15){$\mathbf{k}_{-}  \sim k_{R}$}
\put(9, 40){$+$}
\put(32, 40){$- -$}
\end{picture}\\
\caption{Scale separation with $ \mathbf{k}_{+} \ll \mathbf{k}_{-}, \, k_{R}$.}
\label{fig: Scale separation}
\end{figure}

Let us consider a scale separated system in which the barotropic (+) modes have
wavenumbers much less than those of the baroclinic (-) modes and much less than the Rossby deformation scale,
$ \mathbf{k}_{+} \ll \mathbf{k}_{-}, \,k_{R}$, see figure \ref{fig: Scale separation}. In the most interesting case the wavenumber $ \mathbf{k}_{-} \sim k_{R}$, but we will not use this restriction in our derivation below.

First let us consider the evolution of the small-scale, baroclinic modes (with wave vector $\mathbf{k}$) and their nonlocal interaction with the large-scale zonal flows. To do this, take the baroclinic part of the kinetic equation (\ref{Baroclinic part}) where the wavenumbers $\mathbf{k},1,2$ used for simplicity before will be changed back to $\mathbf{k},\mathbf{k}_{1},\mathbf{k}_{2}$:
\begin{eqnarray}
\partial_{t}n_{\mathbf{k}}^{-}&=&\int W_{\mathbf{k},\mathbf{k}_{1},\mathbf{k}_{2}}^{-+-}[n_{\mathbf{k}_{1}}^{+}n_{\mathbf{k}_{2}}^{-}+n_{\mathbf{k}_{1}}^{+}n_{\mathbf{k}}^{-}sign(\omega_{\mathbf{k}}\omega_{\mathbf{k}_{2}})+n_{\mathbf{k}_{2}}^{-}n_{\mathbf{k}}^{-}sign(\omega_{\mathbf{k}}\omega_{\mathbf{k}_{1}})]\delta(\omega_{\mathbf{k}}^{-}+\omega_{\mathbf{k}_{1}}^{+}+\omega_{\mathbf{k}_{2}}^{-})\nonumber\\
&&\delta(\mathbf{k}+\mathbf{k}_{1}+\mathbf{k}_{2})d\mathbf{k}_{12}.\nonumber
\end{eqnarray}
The third term $n_{\mathbf{k}_{2}}^{-}n_{\mathbf{k}}^{-}$ can be neglected because it is quadratic with respect to the small scales. In the second term $sign(\omega_{\mathbf{k}}\omega_{\mathbf{k}_{2}})\rightarrow -1$ since $\mathbf{k}_{1}$ is the barotropic wavenumber and small so, from the resonance conditions, $\mathbf{k}_{2}\simeq -\mathbf{k}$ and $\omega_{\mathbf{k}_{2}}\simeq -\omega_{\mathbf{k}}.$ Hence equation (\ref{Baroclinic part}) reduces to:
\begin{equation}
\label{Reduced collision integral}
\partial_{t}n_{\mathbf{k}}^{-}=\int W_{\mathbf{k},\mathbf{k}_{1},-\mathbf{k}-\mathbf{k}_{1}}^{-+-}n_{\mathbf{k}_{1}}^{+}[n_{-\mathbf{k}-\mathbf{k}_{1}}^{-}-n_{\mathbf{k}}^{-}]\delta(\omega_{\mathbf{k}}^{-}+\omega_{\mathbf{k}_{1}}^{+}+\omega_{-\mathbf{k}-\mathbf{k}_{1}}^{-})\delta_{\mathbf{k}12}d\mathbf{k}_{1},
\end{equation}
where we integrated out $\mathbf{k}_{2}$ writing it as $-\mathbf{k}-\mathbf{k}_{1}$. Now let:
\begin{equation}
\partial_{t}n_{\mathbf{k}}^{-}=\int F(\mathbf{k},\mathbf{k}_{1})d\mathbf{k}_{1},
\end{equation}
where:
\begin{equation}
\label{F}
F(\mathbf{k},\mathbf{k}_{1})=W_{\mathbf{k},\mathbf{k}_{1},-\mathbf{k}-\mathbf{k}_{1}}^{-+-}n_{\mathbf{k}_{1}}^{+}[n_{-\mathbf{k}-\mathbf{k}_{1}}^{-}-n_{\mathbf{k}}^{-}]\delta(\omega_{\mathbf{k}}^{-}+\omega_{\mathbf{k}_{1}}^{+}+\omega_{-\mathbf{k}-\mathbf{k}_{1}}^{-}).
\end{equation}
Using the symmetries:
$$W_{\mathbf{k},\mathbf{k}_{1},-\mathbf{k}-\mathbf{k}_{1}}^{-+-}=W_{-\mathbf{k}-\mathbf{k}_{1},\mathbf{k}_{1},\mathbf{k}}^{-+-},$$ and: $$W_{\mathbf{k},\mathbf{k}_{1},-\mathbf{k}-\mathbf{k}_{1}}^{-+-}=W_{-\mathbf{k},-\mathbf{k}_{1},\mathbf{k}+\mathbf{k}_{1}}^{-+-},$$
we get:
\begin{equation}
F(\mathbf{k},\mathbf{k}_{1})=-F(-\mathbf{k}-\mathbf{k}_{1},\mathbf{k}_{1})=-F(\mathbf{k}+\mathbf{k}_{1},-\mathbf{k}_{1}),
\end{equation}
so we can write:
\begin{eqnarray}
\partial_{t}n_{\mathbf{k}}^{-}&=&\frac{1}{2}\int (F(\mathbf{k},\mathbf{k}_{1})-F(\mathbf{k}+\mathbf{k}_{1},-\mathbf{k}_{1}))d\mathbf{k}_{1}\\
&=&\frac{1}{2}\int (F(\mathbf{k},\mathbf{k}_{1})-F(\mathbf{k}-\mathbf{k}_{1},\mathbf{k}_{1}))d\mathbf{k}_{1}.\nonumber
\end{eqnarray}
Taylor expand $F(\mathbf{k}-\mathbf{k}_{1},\mathbf{k}_{1})$ with respect to $\mathbf{k}_{1}$ and neglect terms of $O(\mathbf{k}_{1}^{2})$ to get:
\begin{equation}
\label{Taylor expansion 1}
\partial_{t}n_{\mathbf{k}}^{-}=\frac{1}{2}\int \mathbf{k}_{1}.\nabla_{\mathbf{k}}F(\mathbf{k},\mathbf{k}_{1})d\mathbf{k}_{1}.
\end{equation}
We now use Taylor expansion in equation (\ref{F}) and rewrite $F$ as:
\begin{eqnarray}
\label{Taylor expansion 2}
F(\mathbf{k},\mathbf{k}_{1})&\approx&W_{\mathbf{k},\mathbf{k}_{1},\mathbf{k}+\mathbf{k}_{1}}^{-+-}\delta(\omega_{\mathbf{k}}^{-}+\omega_{\mathbf{k}_{1}}^{+}+\omega_{\mathbf{k}+\mathbf{k}_{1}}^{-})(\mathbf{k}_{1}.\nabla_{\mathbf{k}}n_{\mathbf{k}}^{-})n_{\mathbf{k}_{1}}^{+}.
\end{eqnarray}
%where we used the fact that %$n_{\mathbf{k}+\mathbf{k}_{1}}^{-}-n_{\mathbf{k}}^{-}=\mathbf{k}_{1}.\nabla_{\mathbf{k}}n_{\mathbf{k}}^{-}+O(\mathbf{k}_{1}^{2}).$
Combining equation (\ref{Taylor expansion 1}) and (\ref{Taylor expansion 2}) gives:
\begin{equation}
\partial_{t}n_{\mathbf{k}}^{-}=\frac{1}{2}\int \mathbf{k}_{1}.\nabla_{\mathbf{k}}\left(W_{\mathbf{k},\mathbf{k}_{1},\mathbf{k}+\mathbf{k}_{1}}^{-+-}\delta(\omega_{\mathbf{k}}^{-}+\omega_{\mathbf{k}_{1}}^{+}+\omega_{\mathbf{k}+\mathbf{k}_{1}}^{-})(\mathbf{k}_{1}.\nabla_{\mathbf{k}}n_{\mathbf{k}}^{-})n_{\mathbf{k}_{1}}^{+}\right)d\mathbf{k}_{1}.
\end{equation}

Similar to work done in the one-layer case in \citep{connaughtonnazarenkoetal2010} the kinetic equation for the small scales $n_{\mathbf{k}}^{-}$ can be written as the following anisotropic diffusion equation in $\mathbf{k}$-space:
\begin{equation}
\label{Diffusion equation}
\frac{\partial n_{\mathbf{k}}^{-}}{\partial t}=\frac{\partial}{\partial k_{i}}S_{ij}\frac{\partial n_{\mathbf{k}}^{-}}{\partial k_{j}},
\end{equation}
where the diffusion tensor:
\begin{equation}
\label{Diffusion tensor}
S_{ij}=\frac{1}{2}\int W_{\mathbf{k},\mathbf{k}_{1},\mathbf{k}+\mathbf{k}_{1}}^{-+-}\delta(\omega_{\mathbf{k}}^{-}+\omega_{\mathbf{k}_{1}}^{+}+\omega_{\mathbf{k}+\mathbf{k}_{1}}^{-})n_{\mathbf{k}_{1}}^{+}k_{1i}k_{1j}d\mathbf{k}_{1},
\end{equation}
depends on the structure of the large scales, $n_{\mathbf{k}_{1}}^{+}.$ Now let us look at the delta term. We can write out the frequency resonant condition:
\begin{equation}
\omega_{\mathbf{k}}^{-}+\omega_{\mathbf{k}_{1}}^{+}+\omega_{\mathbf{k}+\mathbf{k}_{1}}^{-}=0,
\end{equation} using the dispersion relations as follows:
\begin{equation}
\frac{k_{x}}{F_{-}+k^{2}}+\frac{k_{1x}}{F_{+}+k_{1}^{2}}+\frac{-k_{x}-k_{1x}}{F_{-}+(k+k_{1})^{2}}=0.
\end{equation}
Since $F_{-}=F_{+}\times H^{2}/h_{1}h_{2}$ (see \citep{kozlovrezniketal1987}), $F_{+}\ll F_{-}$ so we can let $F_{+}\rightarrow 0$ and remove it from the above equation. Let us assume that the scaling:
\begin{equation}
\label{Assumption}
k_{1y}^{3}\sim k_{1x},
\end{equation}
is true. Using this assumption we get:
\begin{eqnarray}
\label{Delta term}
&&\frac{k_{x}}{F_{-}+k^{2}}+\frac{k_{1x}}{k_{1y}^{2}}-\frac{k_{x}+k_{1x}}{F_{-}+k^{2}+2k_{x}k_{1x}+2k_{y}k_{1y}+k_{1y}^{2}}\\
&=&\frac{k_{x}}{F_{-}+k^{2}}+\frac{k_{1x}}{k_{1y}^{2}}+\frac{k_{x}}{F_{-}+k^{2}}\left(-1+\frac{2k_{y}k_{1y}}{F_{-}+k^{2}}\right)+O(k_{1y}^{2}).\nonumber
\end{eqnarray}
So from (\ref{Delta term}) we are left with:
\begin{eqnarray}
\label{Delta term 2}
\delta(\omega_{\mathbf{k}}^{-}+\omega_{1}^{+}+\omega_{2}^{-})&=&\delta\left(\frac{k_{1x}}{k_{1y}^{2}}+\frac{2k_{x}k_{y}k_{1y}}{(F_{-}+k^{2})^{2}}\right)\\
&=&k_{1y}^{2}\delta\left(k_{1x}+\frac{2k_{x}k_{y}k_{1y}^{3}}{(F_{-}+k^{2})^{2}}\right).\nonumber
\end{eqnarray}
From which we can see that:
\begin{equation}
\label{k1x}
k_{1x}=-k_{1y}^{3}\frac{2k_{x}k_{y}}{(F_{-}+k^{2})^{2}}.
\end{equation}
Hence:
\begin{equation}
\label{1x<1y}
k_{1x}\ll k_{1y}^{2},
\end{equation}
and the scaling (\ref{Assumption}) is confirmed. From equation (\ref{1x<1y}), terms containing $k_{1x}$ can be removed from the diffusion equation (\ref{Diffusion equation}) leaving:
\begin{equation}
\label{Diffusion equation 2}
\frac{\partial n_{\mathbf{k}}^{-}}{\partial t}=\frac{\partial}{\partial k_{1y}}S_{yy}\frac{\partial n_{\mathbf{k}}^{-}}{\partial k_{1y}},
\end{equation}
which describes diffusion in the $k_{y}$ direction with $k_{x}$ constant. From (\ref{Diffusion tensor}) we can write:
\begin{eqnarray}
S_{yy}&=&\frac{1}{2}\int\limits_{-\infty}^{\infty}W_{\mathbf{k},\mathbf{k}_{1},\mathbf{k}+\mathbf{k}_{1}}^{-+-}k_{1y}^{2}\delta\left(k_{1x}+\frac{2k_{x}k_{y}k_{1y}^{3}}{(F_{-}+k^{2})^{2}}\right)n_{\mathbf{k}_{1}}^{+}k_{1y}^{2}dk_{1x}dk_{1y}\\
&=&\frac{1}{2}\int\limits_{-\infty}^{\infty}W_{\mathbf{k},\mathbf{k}_{1},\mathbf{k}+\mathbf{k}_{1}}^{-+-}\delta(k_{1x}+\theta k_{1y}^{3})n_{\mathbf{k}_{1}}^{+}k_{1y}^{4}dk_{1x}dk_{1y},
\end{eqnarray}
where $\theta=\frac{2k_{x}k_{y}}{(F_{-}+k^{2})^{2}}.$ Since $k_{1x}=-\theta k_{1y}^{3},$ from equation(\ref{k1x}), we get:
\begin{equation}
\label{Syy}
S_{yy}=\frac{1}{2}\int\limits_{-\infty}^{\infty}\left[W_{\mathbf{k},\mathbf{k}_{1},\mathbf{k}+\mathbf{k}_{1}}^{-+-}n_{\mathbf{k}_{1}}^{+}\right]_{k_{1x}=-\theta k_{1y}^{3}}k_{1y}^{4}dk_{1y}.
\end{equation}

To close the system, baroclinic equation (\ref{Diffusion equation 2}) has to be complemented by a barotropic equation, which is obtained from equation (\ref{Barotropic part}) in which the term $n_{\mathbf{k}_{1}}^{-}n_{\mathbf{k}_{2}}^{-}$ is neglected (because $n_{\mathbf{k}}^{+}\gg n_{\mathbf{k}_{2}}^{-}$). This gives:
\begin{equation}
\label{Barotropic reduced}
\partial_{t}n_{\mathbf{k}}^{+}=2\int W_{\mathbf{k},\mathbf{k}_{1},\mathbf{k}_{2}}^{+--}n_{\mathbf{k}_{1}}^{-}n_{\mathbf{k}}^{+}sign(k_{x}k_{2x})\delta(\omega_{\mathbf{k}}^{+}+\omega_{\mathbf{k}_{1}}^{-}+\omega_{\mathbf{k}_{2}}^{-})\delta_{\mathbf{k}12}d\mathbf{k}_{12}.
\end{equation}

\begin{figure}[h!]
\centering
\setlength{\unitlength}{0.5mm}
\begin{picture}(150, 300)(50, -20)
\put(0, 125){\line(1, 0){250}} % x-axis
\put(125, 0){\line(0, 1){250}} % y-axis
\put(125, 125){\circle{32}} % small circle
\bigcircle{125}{125}{90} % big inside circle
\bigcircle{125}{125}{110} % big outside circle
\put(0, 220){Small scales}
\put(20, 210){\vector(1, -1){25}}
\put(75, 90){Large scales}
\put(95, 100){\vector(1, 1){22}}
\put(170, 215){\vector(0, 1){40}} % wave-action top
\put(190, 200){\vector(0, 1){40}}
\put(210, 185){\vector(0, 1){40}}
\put(180, 250){Wave-action}
\put(170, 35){\vector(0, -1){40}} % wave-action bottom
\put(190, 50){\vector(0, -1){40}}
\put(210, 65){\vector(0, -1){40}}
\put(180, 0){Wave-action}
\put(225, 122){\line(0, 1){5}} % deformation scale
\put(220, 110){$\mathbf{k}\sim (k_{R},0)$}
\put(255, 122){$k_{x}$}
\put(122, 255){$k_{y}$}
\qbezier(200, 190)(160, 180)(140, 140) % energy curve
\put(140, 140){\vector(-1, -2){1}} % energy arrow head
\put(140, 175){Energy}
\qbezier(120, 138)(129, 125)(120, 112) % parabola
\qbezier(130, 138)(121, 125)(130, 112)
\put(150, 90){$k_{x}\sim k_{y}^{3}$}
\put(148, 95){\vector(-1, 1){20}}
\end{picture}\\
\caption{Diagram to show the direction of wave-action and energy transfer.}
\label{fig: Transfer directions}
\end{figure}
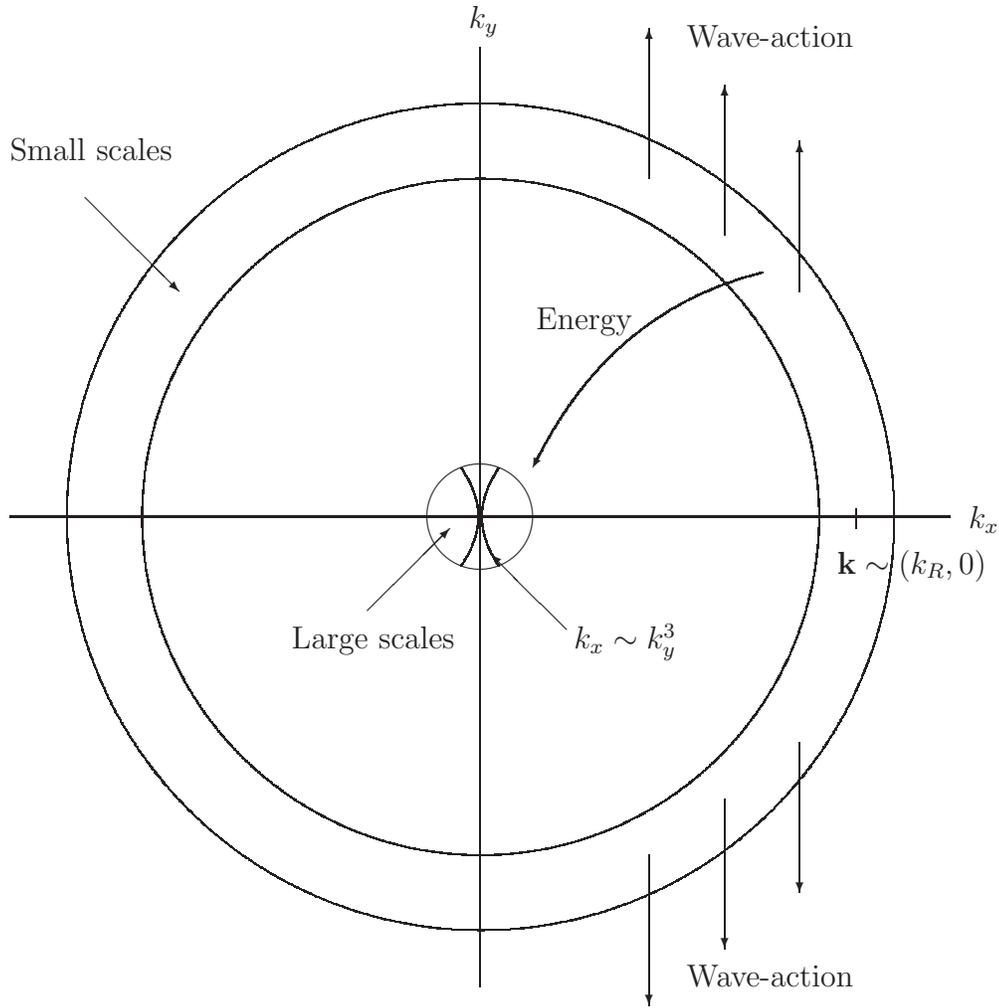

Thus we obtained the system of coupled equations (\ref{Diffusion equation 2}) and (\ref{Barotropic reduced}) for the small-scale baroclinic component and the large-scale barotropic component. One can see that the total wave-action is conserved in the small-scale baroclinic component alone. This is natural because the nonlocal process that we considered can be interpreted as scattering of small-scale baroclinic wave packets of a slowly varying barotropic flow. The number of wave packets in such a process in conserved. On the other hand the total energy in the small-scale component is not conserved. Only the sum of the energies of the small-scale baroclinic and the large-scale barotropic components are conserved (see appendix \ref{app: Conservation}). Thus the energy may be exchanged between the small-scale baroclinic and large-scale barotropic components. The dominant transfer direction is from small to large scales. Indeed consider an initial small-scale spectrum that is concentrated near the meridional axis with $\mathbf{k}\sim (k_{x0},0)$. According to our diffusion equation (\ref{Diffusion equation 2}) this spectrum will spread in $k_{y}$ which means that the frequency will spread towards larger $k_{y}$'s with $k_{x}$ remaining fixed. As $k_{y}$ increases the frequency of the respective modes $\omega_{\mathbf{k}}^{-}=k_{x}/(F^{-}+k_{x}^{2}+k_{y}^{2})$ decreases. Considering that the total wave-action is conserved this means that the total small-scale baroclinic energy $\int \omega_{\mathbf{k}}^{-} n_{\mathbf{k}}^{-} \, d\mathbf{k}$ will decrease. Since the total barotropic plus baroclinic energy is conserved the baroclinic energy loss will be compensated by the growth of the barotropic energy at large scales. As the barotropic waves that interact with the small-scale baroclinic modes are mostly zonal (equation (\ref{Assumption})), this transfer of energy will be mostly anisotropic and mostly to the large-scale zonal component. See diagram (\ref{fig: Transfer directions}).

\section{Conclusion}\label{sec: Summary}
Two-layer models are more realistic when describing the oceans, as less dense, warmer surface waters float on top of denser, colder waters. Up until now, unlike the one-layer case, not much work has been done for two layers in a WT context and the work that as been done makes major assumptions such as equal layers, when in fact the top layer is typically much thinner than the bottom layer.

In this paper we began by deriving a symmetric form of the two-layer kinetic equation using canonical variables. We considered an ocean with a thin upper layer and as a result the kinetic equation contained only $\{--+\}$ triads (whose wavenumbers and frequencies are in resonance) with two baroclinic components and one barotropic component. We then studied the turbulent cascade of energy between the barotropic and baroclinic modes. We showed that energy is transferred via local triad interactions from large-scale baroclinic modes to baroclinic and barotropic modes at the Rossby deformation scale. From there, energy is transferred to the large-scale barotropic modes via an inverse nonlocal transfer.
% How are large scales generated directly from small scales? Modulation instability?

Via scale separation a system of coupled equations were obtained for the small-scale baroclinic component and the large-scale barotropic component. From our diffusion equation, it can be seen that diffusion occurs towards larger $k_{y}$'s with $k_{x}$ fixed. Small-scale wave-action $n_{\mathbf{k}}^{-}$ is conserved by this motion.

The energy of the small scales $\omega_{\mathbf{k}}^{-}n_{\mathbf{k}}^{-}$ is not conserved, however. Since the total energy (barotropic plus baroclinic) of the large and small scales together is conserved, baroclinic energy lost by small scales will be compensated by the growth of the barotropic energy at large scales. We find that this transfer is mostly anisotropic and mostly to the zonal component.

It is possible that a negative feedback loop forms, similar to in the one-layer model \citep{connaughtonnazarenkoetal2011}, whereby the growth of large scales should turn off the energy source (which we showed in the case of WT cannot be BI).
%What is it?
One can answer this by direct numerical simulation (DNS) of the two-layer model and this is something to consider in future work.

\section*{Acknowledgements}

S.V. Nazarenko and S.B. Medvedev acknowledge support from the government of Russian Federation via grant No. 12.740.11.1430 for supporting research of teams working under supervision of invited scientists.

\appendix

\section{Conservation of energy and potential enstrophy}\label{app: Conservation}

Let us prove that the two-layer kinetic equation conserves the total energy (of the barotropic and baroclinic modes):
\begin{equation}
\label{Energy}
E=\sum\limits_{\sigma}\int |\omega_{\mathbf{k}}^{\sigma}|n_{\mathbf{k}}^{\sigma}d\mathbf{k},
\end{equation}
and the total potential enstrophy:
\begin{equation}
\label{Enstrophy}
\Omega=\sum\limits_{\sigma}\int \omega_{\mathbf{k}}^{\sigma}(k^{2}+F^{\sigma})n_{\mathbf{k}}^{\sigma}d\mathbf{k}=\sum\limits_{\sigma}\int |k_{x}|n_{\mathbf{k}}^{\sigma}d\mathbf{k},
\end{equation}
by substituting equation (\ref{Symmetric 2-layer kinetic equation}):
\begin{equation}
\dot{n}_{\mathbf{k}}^{\sigma}=\sum\limits_{\mu \nu}\int W_{\mathbf{k}12}^{\sigma \mu \nu}[n_{1}^{\mu}n_{2}^{\nu}+n_{\mathbf{k}}^{\sigma}n_{1}^{\mu}\text{sign}(\omega_{\mathbf{k}}\omega_{2})+n_{\mathbf{k}}^{\sigma}n_{2}^{\nu}\text{sign}(\omega_{\mathbf{k}}\omega_{1})]\delta(\omega_{\mathbf{k}12}^{\sigma \mu \nu})\delta_{\mathbf{k}12}d\mathbf{k}_{12},
\end{equation}
into (\ref{Energy}) and (\ref{Enstrophy}).
\begin{eqnarray}
\dot{E}&=&\sum\limits_{\sigma}\int |\omega_{\mathbf{k}}^{\sigma}|\dot{n}_{\mathbf{k}}^{\sigma}d\mathbf{k}\\
&=&\sum\limits_{\sigma \mu \nu}\int \omega_{\mathbf{k}}^{\sigma}\text{sign}(\omega_{\mathbf{k}})W_{\mathbf{k}12}^{\sigma \mu \nu}[n_{1}^{\mu}n_{2}^{\nu}+n_{\mathbf{k}}^{\sigma}n_{1}^{\mu}\text{sign}(\omega_{\mathbf{k}}\omega_{2})+n_{\mathbf{k}}^{\sigma}n_{2}^{\nu}\text{sign}(\omega_{\mathbf{k}}\omega_{1})]\delta(\omega_{\mathbf{k}12}^{\sigma \mu \nu})\delta_{\mathbf{k}12}d\mathbf{k}_{12}.\nonumber
\end{eqnarray}
Exchanging $\mathbf{k}\leftrightarrow \mathbf{k}_{3}$ we have:
\begin{equation}
\sum\limits_{\sigma \mu \nu}\int W_{312}^{\sigma \mu \nu}[n_{1}^{\mu}n_{2}^{\nu}\omega_{3}^{\sigma}\text{sign}(\omega_{3})+n_{3}^{\sigma}n_{1}^{\mu}\omega_{3}^{\sigma}\text{sign}(\omega_{2})+n_{3}^{\sigma}n_{2}^{\nu}\omega_{3}^{\sigma}\text{sign}(\omega_{1})]
\delta(\omega_{312}^{\sigma \mu \nu})\delta_{312}d\mathbf{k}_{123},
\end{equation}
then swapping $3\leftrightarrow 2$ in the second term and $3\leftrightarrow 1$ in the third term gives:
\begin{equation}
\frac{1}{3}\int W_{312}^{\sigma \mu \nu}n_{1}^{\mu}n_{2}^{\nu}\text{sign}(\omega_{3})(\omega_{3}^{\sigma}+\omega_{2}^{\nu}+\omega_{1}^{\mu})\delta(\omega_{312}^{\sigma \mu \nu})\delta_{312}d\mathbf{k}_{123},
\end{equation}
which is zero by the frequency resonance condition. Similarly for the potential enstrophy:
\begin{eqnarray}
\dot{\Omega}&=&\sum\limits_{\sigma}\int |k_{x}|\dot{n}_{\mathbf{k}}^{\sigma}d\mathbf{k}\\
&=&\sum\limits_{\sigma \mu \nu}\int k_{x}\text{sign}(k_{x})W_{\mathbf{k}12}^{\sigma \mu \nu}[n_{1}^{\mu}n_{2}^{\nu}+n_{\mathbf{k}}^{\sigma}n_{1}^{\mu}\text{sign}(k_{x}k_{2x})+n_{\mathbf{k}}^{\sigma}n_{2}^{\nu}\text{sign}(k_{x}k_{1x})]\delta(\omega_{\mathbf{k}12}^{\sigma \mu \nu})\delta_{\mathbf{k}12}d\mathbf{k}_{12}\nonumber\\
&=&\sum\limits_{\sigma \mu \nu}\int W_{312}^{\sigma \mu \nu}[n_{1}^{\mu}n_{2}^{\nu}k_{3x}\text{sign}(k_{3x})+n_{3}^{\sigma}n_{1}^{\mu}k_{3x}\text{sign}(k_{2x})+n_{3}^{\sigma}n_{2}^{\nu}k_{3x}\text{sign}(k_{1x})]\delta(\omega_{312}^{\sigma \mu \nu})\delta_{312}d\mathbf{k}_{312}\nonumber\\
&=&\sum\limits_{\sigma \mu \nu}\int W_{312}^{\sigma \mu \nu}n_{1}^{\mu}n_{2}^{\nu}\text{sign}(k_{3x})(k_{3x}+k_{2x}+k_{1x})\delta(\omega_{312}^{\sigma \mu \nu})\delta_{312}d\mathbf{k}_{312},\nonumber
\end{eqnarray}
which is zero by the wavenumber resonance condition.

\begin{comment}

\end{comment}

\bibliography{2_layer_bibliography}

\begin{thebibliography}{19}
\providecommand{\natexlab}[1]{#1}
\providecommand{\url}[1]{\texttt{#1}}
\expandafter\ifx\csname urlstyle\endcsname\relax
  \providecommand{\doi}[1]{doi: #1}\else
  \providecommand{\doi}{doi: \begingroup \urlstyle{rm}\Url}\fi

\bibitem[Balk et~al.(1990)Balk, Nazarenko, and Zakharov]{balknazarenkoetal1990}
A.~Balk, S.~Nazarenko, and V.~Zakharov.
\newblock On the nonlocal turbulence of drift type waves.
\newblock \emph{Phys. Lett. A.}, 146:\penalty0 217--221, 1990.

\bibitem[Balk et~al.(1991)Balk, Nazarenko, and Zakharov]{balknazarenkoetal1991}
A.~Balk, S.~Nazarenko, and V.~Zakharov.
\newblock A new invariant for drift turbulence.
\newblock \emph{Phys. Lett. A}, 152:\penalty0 276--280, 1991.

\bibitem[Bartello(1995)]{bartello1995}
P.~Bartello.
\newblock Geostrophic adjustment and inverse cascades in rotating stratifed
  turbulence.
\newblock \emph{J. Atmos. Sci.}, 52:\penalty0 4410--4428, 1995.

\bibitem[Bedard et~al.(2013)Bedard, Lukaschuk, and
  Nazarenko]{bedardlukaschuketal2013}
R.~Bedard, S.~Lukaschuk, and S.~Nazarenko.
\newblock Non-stationary regimes of surface gravity wave turbulence.
\newblock \emph{JETP Lett.}, 97:\penalty0 529--535, 2013.

\bibitem[Charney(1948)]{charney1948}
J.~Charney.
\newblock On the scale of atmospheric motions.
\newblock \emph{Geofys. Publikasjoner}, 17:\penalty0 1--17, 1948.

\bibitem[Connaughton et~al.(2010)Connaughton, Nazarenko, and
  Quinn]{connaughtonnazarenkoetal2010}
C.~Connaughton, S.~Nazarenko, and B.~Quinn.
\newblock Nonlocal wave turbulence in the {Charney}-{Hasegawa}-{Mima} equation:
  a short review.
\newblock \emph{arXiv:1012.2714}, pages~--, 2010.

\bibitem[Connaughton et~al.(2011)Connaughton, Nazarenko, and
  Quinn]{connaughtonnazarenkoetal2011}
C.~Connaughton, S.~Nazarenko, and B.~Quinn.
\newblock Feedback of zonal flows on wave turbulence driven by small-scale
  instability in the {Charney}-{Hasegawa}-{Mima} model.
\newblock \emph{EPL}, 96:\penalty0 25001, 2011.

\bibitem[Janssen(2008)]{janssen2008}
P.~Janssen.
\newblock Progress in ocean wave forecasting.
\newblock \emph{J. Comput. Phys.}, 227:\penalty0 3572--3594, 2008.

\bibitem[Kozlov et~al.(1987)Kozlov, Reznik, and Soomere]{kozlovrezniketal1987}
O.~Kozlov, G.~Reznik, and T.~Soomere.
\newblock Kinetic equation for {Rossby} waves in two-layer ocean.
\newblock \emph{Izv. Akad. Nauk SSSR Ser. Fiz. Atmosfer. i Okeana},
  23:\penalty0 1165--1173, 1987.

\bibitem[Kraichnan(1967)]{kraichnan1967}
R.~Kraichnan.
\newblock Inertial ranges in two-dimensional turbulence.
\newblock \emph{Phys. Fluids}, 10:\penalty0 1417--1423, 1967.

\bibitem[Lin(1980)]{lin1980}
C.~Lin.
\newblock Eddy heat fluxes and stability of planetary waves: part {I}.
\newblock \emph{J. Atm. Sci.}, 37:\penalty0 2353--2372, 1980.

\bibitem[L'vov et~al.(2006)L'vov, Nazarenko, and Skrbek]{lvovnazarenkoetal2006}
V.~L'vov, S.~Nazarenko, and L.~Skrbek.
\newblock Energy spectra of developed turbulence in helium superfluids.
\newblock \emph{JLTP}, 145:\penalty0 125--142, 2006.

\bibitem[McWilliams(2006)]{mcwilliams2006}
J.~McWilliams.
\newblock \emph{Fundamentals of Geophysical Fluid Dynamics}.
\newblock Cambridge University Press, 2006.

\bibitem[Nazarenko(2011)]{nazarenko2011}
S.~Nazarenko.
\newblock \emph{Wave Turbulence (Lecture notes in Physics 825)}.
\newblock Springer, 2011.

\bibitem[Nazarenko and Quinn(2009)]{nazarenkoquinn2009}
S.~Nazarenko and B.~Quinn.
\newblock Triple cascade behaviour in {QG} and drift turbulence and the
  generation of zonal jets.
\newblock \emph{Phys. Rev. Lett.}, 103:\penalty0 118501, 2009.

\bibitem[Phillips(1951)]{phillips1951}
N.~Phillips.
\newblock A simple three-dimensional model for the study of large-scale
  extra-tropical flow patterns.
\newblock \emph{J. Meteor}, 8:\penalty0 381--394, 1951.

\bibitem[Salmon(1978)]{salmon1978}
R.~Salmon.
\newblock Two-layer quasi-geostrophic turbulence in a simple special case.
\newblock \emph{Geophys. Astrophys. Fluid Dynamics}, 10:\penalty0 25--52, 1978.

\bibitem[Tronko et~al.(2013)Tronko, Nazarenko, and
  Galtier]{tronkonazarenkoetal2013}
N.~Tronko, S.~Nazarenko, and S.~Galtier.
\newblock Weak turbulence in two-dimensional magnetohydrodynamics.
\newblock \emph{Phys. Rev. E}, 87:\penalty0 033103, 2013.

\bibitem[Zakharov et~al.(1992)Zakharov, L'vov, and
  Falkovich]{zakharovlvovetal1992}
V.~Zakharov, V.~L'vov, and G.~Falkovich.
\newblock \emph{{Kolmogorov} spectra of turbulence {I}: {Wave} turbulence}.
\newblock Springer, 1992.

\end{thebibliography}


\begin{thebibliography}{21}
\bibitem{1} Charney J G (1948) On the scale of atmospheric motions {\it {Geofys. Publikasjoner}} {\bf{17}} pp~1--17
\bibitem{2} Phillips N A (1951) A simple three-dimensional model for the study of large-scale extra-tropical flow patterns {\it {J. Meteor}} {\bf{8}} pp~381--394
\bibitem{3} Kraichnan R H (1967) Inertial ranges in two-dimensional turbulence {\it {Phys. Fluids}} {\bf{10}} pp~1417--1423
\bibitem{4} Salmon R (1978) Two-layer quasi-geostrophic turbulence in a simple special case {\it {Geophys. Astrophys. Fluid Dynamics}} {\bf{10}} pp~25--52
\bibitem{5} Zakharov V E, L'vov V S and Falkovich G (1992) Statistical description of weak wave turbulence {\it{Kolmogorov spectra of turbulence I}} (Springer Berlin Heidelberg) pp~63--82
\bibitem{6} Nazarenko S V (2011) Wave turbulence {\it{Springer-Verlag: Berlin Heidelberg}}
\bibitem{7} Lvov V S, Nazarenko S V and Skrbek L (2006) Energy spectra of developed turbulence in helium superfluids {\it{JLTP}} {\bf{145}} pp~125--142
\bibitem{8} Bedard R, Lukaschuk S and Nazarenko S V (2013) Non-stationary regimes of surface gravity wave turbulence {\it{JETP Letter}} 97 (7-8)
\bibitem{9} Tronko N, Nazarenko S V and Galtier S (2013) Weak turbulence in two-dimensional magnetohydrodynamics {\it{Physical Review E}} 87 (3)
\bibitem{10} Bartello P (1995) Geostrophic adjustment and inverse cascades in rotating stratified turbulence {\it{J. Atmos. Sci.}} {\bf{52}} pp~4410–-4428
\bibitem{11} Zakharov V E and Filonenko N N (1966) The energy spectrum for stochastic oscillations of a fluid surface {\it{Doclady Akad. Nauk SSSR}} {\bf{170}} pp~1292--1295 [Sov. Phys. Docl. 11, 881-884 (1967)]
\bibitem{12} Janssen P (2008) Progress in ocean wave forecasting {\it{Journal of Computational Physics}} {\bf{227}} pp~3572–-3594
\bibitem{13} Kozlov O V, Reznik G M and Soomere T E (1987) Kinetic equation for Rossby waves in two-layer ocean {\it{Izv. Akad. Nauk SSSR Ser. Fiz. Atmosfer. i Okeana}} {\bf{23}} pp~1165--1173
\bibitem{14} McWilliams J C (2006) Fundamentals of geophysical fluid dynamics {\it{Cambridge University Press: Cambridge}} pp~178--184
\bibitem{15} Balk A M, Nazarenko S V and Zakharov V E (1991) A New Invariant for Drift Turbulence {\it{Phys. Lett. A}} {\bf{152}} pp~276--280
\bibitem{16} Nazarenko S V and Quinn B (2009) Triple cascade behaviour in QG and drift turbulence and generation of zonal jets {\it{Physical Review Letters}} {\bf{103}}
\bibitem{17} Balk A M, Nazarenko S V and Zakharov V E (1990) On the nonlocal turbulence of drift type waves {\it{Letters A}} {\bf{146}} pp~217–-221
\bibitem{18} Balk A M, Nazarenko S V and Zakharov V E (1990) Nonlocal turbulence of drift waves {\it{Zh. Eksp. Teor. Fiz.}} {\bf{98}} pp~446--467
\bibitem{19} Connaughton C, Nazarenko S V and Quinn B (2010) Nonlocal wave turbulence in the Charney-Hasegawa-Mima equation: a short review arXiv:1012.2714
\bibitem{20} Connaughton C, Nazarenko S V and Quinn B (2011) Feedback of zonal flows on wave turbulence driven by small-scale instability in the Charney-Hasegawa-Mima model {\it{EPL}} {\bf{96}}
\bibitem{21} Lin C.~A. (1980) Eddy heat fluxes and stability of planetary waves. Part I {\it{J. Atm. Sci.}} {\bf{37}} pp~2353--2372
\bibitem{} Salmon R (1980) Baroclinic instability and geostrophic turbulence {\it{Geophys. Astrophys. Fluid Dynamics}} {\bf{15}} pp~167--211
\bibitem{} Soomere T (1996) Spectral evolution of two-layer weak geostrophic turbulence. Part 1: Typical scenarios {\it{Nonlinear Processes in Geophysics}} {\bf{3}} pp~166--195
\end{thebibliography}

\end{document}